\documentclass[useAMS,useAMSmath, usenatbib, usegraphicx, referee]{mn2e}

\usepackage{amsbsy}

\title[Electron-positron energy deposition near neutron and quark stars]{Electron-positron energy deposition rate from neutrino pair annihilation in the equatorial plane of rapidly rotating neutron and quark stars}
\author[Kov\'{a}cs, Cheng and Harko]{Z. Kov\'{a}cs\thanks{E-mail:
zkovacs@mpifr-bonn.mpg.de}, K. S. Cheng\thanks{E-mail:
hrspksc@hkucc.hku.hk} and T. Harko\thanks{E-mail:
harko@hkucc.hku.hk}
\\
Department of Physics and Center for Theoretical and Computational Physics, The University of Hong Kong, \\
Pok Fu Lam Road, Hong Kong, Hong Kong SAR, P. R. China
}
\begin{document}


\pagerange{\pageref{firstpage}--\pageref{lastpage}} \pubyear{2002}

\maketitle


\def\apj{Astrophys. J.}
\def\mnras{Month. Not. Roy. Astr. Soc.}
\def\prd{Phys. Rev. D}

\date{Accepted 1988 December 15. Received 1988 December 14; in original form 1988 October 11}

\pagerange{\pageref{firstpage}--\pageref{lastpage}} \pubyear{2002}

\maketitle

\label{firstpage}

\begin{abstract}
The neutrino-antineutrino annihilation into electron-positron pairs near the surface of compact general relativistic  stars could play an important role in supernova explosions, neutron star collapse, or for close neutron star binaries near their last stable orbit. General relativistic effects increase the energy deposition rates due to the annihilation process. We investigate the deposition of energy and momentum due to the annihilations of neutrinos and antineutrinos in the equatorial plane of the rapidly rotating neutron and quark stars, respectively. We analyze the influence of general relativistic effects, and we obtain the general relativistic corrections to the energy and momentum  deposition rates for arbitrary stationary and axisymmetric space-times. We obtain the energy and momentum deposition rates for several classes of rapidly rotating neutron stars, described by different equations of state of the neutron matter, and for quark stars, described by the MIT bag model equation of state and in the CFL (Color-Flavor-Locked) phase, respectively. Compared to the Newtonian calculations, rotation and general relativistic effects increase the total annihilation rate measured by an observer at infinity. The differences in the equations of state for neutron and quark matter also have important effects on the spatial distribution of the energy deposition rate by neutrino-antineutrino annihilation.
\end{abstract}

\begin{keywords}
neutrinos: dense matter -- equation of state: stars: rotation: relativity.
\end{keywords}

\section{Introduction}

Since the pioneering works of Cooperstein et al. (1986), Cooperstein et al. (1987),  and Goodman et al. (1987), the energy deposition rate from the $\nu +\bar{\nu}\rightarrow e^{+}+e^{-}$ neutrino annihilation reaction has been intensively studied.  This reaction is of considerable importance for Type II supernova dynamics, neutron star collapse, or for close neutron star binaries near their last stable orbit. Neutrino-antineutrino annihilation into electrons and positrons can deposit more than $10^{51}$ ergs above the neutrino-sphere of a type II supernova \citep{Go87}. This energy deposition, together with neutrino-baryon capture, significantly increases the neutrino heating in the envelope via the so-called delayed shock mechanism \citep{BeWi85,Be90}. For large $r$ the energy deposition rate is proportional to $r^{-8}$, where $r$ is the distance from the center of the neutrino-sphere. The initial estimations of the neutrino annihilation reaction efficiencies were based on a Newtonian approach, by assuming that $2GM/c^2R<<1$, where $M$ is the gravitational mass of the star, and $R$ is the distance scale. However, for a full understanding of the effects of the neutrino annihilation in strong gravitational fields, general relativistic effects must be taken into account \citep{SaWi99}. For a static neutron star, by adopting for the description of the gravitational field the Schwarzschild metric,  the efficiency of the $\nu +\bar{\nu}\rightarrow e^{+}+e^{-}$ process is enhanced over the Newtonian values up to a factor of more than $4$ in the regime applicable to Type II supernovae, and by up to a factor of $30$ for collapsing neutron stars \citep{SaWi99}. The neutrino pair annihilation rate into electron pairs between two neutron stars in a binary system was calculated by Salmonson \& Wilson (2001). A closed formula for the energy deposition rate at any point between the stars was obtained, where each neutrino of a pair derives from each star, and this result was compared with that in which all neutrinos derive from a single neutron star. An approximate generalization of this formula was also given to include the relativistic effects of gravity. The interstar neutrino annihilation is a significant contributor to the energy deposition between heated neutron star binaries.

The neutrino-antineutrino annihilation into electrons and positrons is an important candidate to explain the energy source of the gamma ray bursts (GRBs) \citep{Pa90, MeRe92,RuJa98, RuJa99, AsIw02}. The semi-analytical study of  the gravitational effects on neutrino pair annihilation near the neutrinosphere and around the thin accretion disk were considered in Asano \& Fukuyama (2000), by assuming that the accretion disk is isothermal, and that the gravitational field is dominated by the Schwarzschild black hole. General relativistic effects were studied only near the rotation axis. The energy deposition rate is enhanced by the effect of orbital bending toward the center. However, the effects of the redshift and gravitational trapping of the deposited energy reduce the effective energy of the gamma-ray burst's source. Although each effect is substantial, the effects partly cancel one another. As a result, the gravitational effects do not substantially change the energy deposition rate for either the spherically symmetric case or the disk case \citep{AsFu00}. Using idealized models of the accretion disk, Asano \& Fukuyama (2001) investigated the relativistic effects on the energy deposition rate via neutrino pair annihilation near the rotation axis of a Kerr black hole, by assuming that the neutrinos are emitted from the accretion disk. The bending of neutrino trajectories and the redshift due to the disk rotation and gravitation were also taken into consideration. The Kerr parameter, $a$, affects not only behavior of the neutrinos, but also the inner radius of the accretion disk. When the deposition energy is mainly contributed by the neutrinos coming from the central part, the redshift effect becomes dominant as a becomes large, and the energy deposition rate is reduced compared with that neglecting the relativistic effects. On the other hand, for a small $a$, the bending effect becomes dominant and makes the energy increase by factor of 2, compared with that which neglects the relativistic effects \citep{AsFu01}.

The effect of the inclusion of the slow rotation of the star in the general relativistic treatment of
the neutrino-antineutrino annihilation into electron positron pairs was considered in Prasanna \& Goswami (2002).
It was shown that the inclusion of the rotation results in a reduction in the heating rate, as compared to the no rotation case. The energy-momentum deposition rate (MDR) from the $\nu -\bar{\nu}$ collisions above a rotating black hole/thin accretion disk system was calculated by Miller et al. (2003), by imaging the accretion disk at a specified observer using the full geodesic equations, and calculating the cumulative MDR from the scattering of all pairs of neutrinos and antineutrinos arriving at the observer. The dominant contribution to the MDR comes from near the surface of the disk with a tilt of approximately $\pi/4$ in the direction of the disk's rotation. The MDR at large radii is directed outward in a conic section centered around the symmetry axis and is larger by a factor of 10-20 than the on-axis values. There is also a linear dependence of the MDR on the black hole angular momentum.
The deposition of energy and momentum, due to the annihilation of neutrinos $\nu $ and
antineutrinos $\bar{\nu }$ in the vicinity of steady, axisymmetric accretion tori around stellar-mass black holes was investigated in Birkl et al. (2007). The influence of general relativistic effects were analyzed in combination
with different neutrinosphere properties and spatial distribution of the
energy deposition rate. Assuming axial symmetry, the annihilation rate 4-vector was numerically computed.
The local neutrino distribution was constructed by ray-tracing neutrino trajectories in a Kerr
space-time, using null geodesics. Different shapes of the neutrinospheres,
spheres, thin disks, and thick accretion tori were studied, whose structure ranges from idealized tori to equilibrium non-selfgravitating matter distributions. Compared to Newtonian calculations,
general relativistic effects increase the total annihilation rate measured by an observer at infinity by a factor of two when the
neutrinosphere is a thin disk, but the increase is only 25\% for toroidal and spherical neutrinospheres. Thin disk models yield the highest energy deposition rates for neutrino-antineutrino annihilation, and spherical neutrinospheres the lowest ones, independently of whether general relativistic effects are included. General relativity and rotation cause important differences in the spatial distribution of the energy deposition rate by neutrino $\nu $ and antineutrino $\bar{\nu }$-annihilation \citep{Bi07}. The study of the structure of neutron star disks based on the two-region (i.e., inner and outer) disk scenario was performed by Zhang \& Dai (2009), who calculated the neutrino annihilation luminosity from the disk in various cases. The effects of the viscosity parameter $\alpha $, energy parameter $\epsilon $ (measuring the neutrino cooling efficiency of the inner disk), and outflow strength on the structure of the entire disk, as well as the effect of emission from the neutron star surface boundary emission on the total neutrino annihilation rate were investigated.  An outflow from the disk decreases the density and pressure, but increases the thickness of the disk. Moreover, compared with the black hole disk, the neutrino annihilation luminosity above the neutron star disk is higher, and the neutrino emission from the boundary layer could increase the neutrino annihilation luminosity by about one order of magnitude higher than the disk without boundary emission. The neutron star disk with the advection-dominated inner disk could produce the highest neutrino luminosity, while the disk with an outflow has the lowest\citep{ZhDa09}. A detailed general relativistic calculation of the neutrino path for a general metric describing a rotating star was studied in Mallick \& Majumder (2009). The neutrino path was calculated along the equatorial and polar plane. The expression for the minimum photosphere radius  was obtained and matched with the Schwarzschild limit. The minimum photosphere radius was calculated for stars with two different equations of state, each rotating with two different velocities. The results show that the minimum photosphere radius for the hadronic star is much greater than for the quark star, and that the minimum photosphere radius increases as the rotational velocity of the star decreases. The minimum photosphere radius along the polar plane is larger than that along the equatorial plane. The estimate of the energy deposition rate for neutrino pair annihilation for the neutrinos coming from the equatorial plane of a rotating neutron star was calculated along the rotation axis in Bhattacharyya et al. (2009), by using the Cook-Shapiro-Teukolsky metric. The neutrino trajectories, and hence the neutrino emitted from the disk, are affected by the redshift due to the disk rotation and gravitation. The energy deposition rate is very sensitive to the value of the temperature, and its variation along the disk. The rotation of the star has a negative effect on the energy deposition rate, it decreases with increase in rotational velocity. The standard model for Type II supernovae explosions, confirmed by the detection of neutrinos emitted during the supernova explosion, predicts the formation of a compact object, usually assumed to be a neutron star \citep{Cha09}. However, the newly formed neutron star at the center of SN 1987A may undergo a phase transition after the neutrino trapping timescale ($\sim10$ s). Consequently the compact remnant of SN 1987A may be a strange quark star, which has a softer equation of state than that of neutron star matter \citep{Cha09}. Such a phase transition can induce stellar collapse and result in large amplitude stellar oscillations. A three-dimensional Newtonian hydrodynamic code was used to study the time evolution of the temperature and density at the neutrinosphere. Extremely intense pulsating neutrino fluxes, with submillisecond period and with neutrino energy (greater than 30 MeV), can be emitted because the oscillations of the temperature and density are out of phase almost $180^{\circ}$. The dynamical evolution of a phase-transition-induced collapse of a neutron star to a hybrid star, which consists of a mixture of hadronic matter and strange quark matter, was studied in Cheng et al. (2009). It was found that both the temperature and the density at the neutrinosphere are oscillating with acoustic frequency. Consequently, extremely intense, pulsating neutrino/antineutrino fluxes will be emitted periodically. Since the energy and density of neutrinos at the peaks of the pulsating fluxes are much higher than the non-oscillating case, the electron/positron pair creation rate can be enhanced dramatically. Some mass layers on the stellar surface can be ejected, by absorbing energy of neutrinos and pairs. These mass ejecta can be further accelerated to relativistic speeds by absorbing electron/positron pairs, created by the neutrino and antineutrino annihilation outside the stellar surface.

It is the purpose of the present paper to consider a comparative systematic study of the neutrino-antineutrino annihilation process around rapidly rotating neutron and strange stars, respectively, and to obtain the basic physical parameters characterizing this process (the electron-positron energy deposition rate per unit volume and unit time, and the total emitted power, respectively), by taking into account the full general relativistic corrections. In order to obtain the electron-positron energy deposition rate for various types of neutron and quark stars we generalize the relativistic description of the neutrino-antineutrino annihilation process to the case of arbitrary stationary and axisymmetric geometries.   To compute the electron-positron energy deposition rate, the metric outside the rotating general relativistic stars must be determined. In the present study we study the equilibrium configurations of the rotating neutron and quark stars by using the RNS code, as introduced in \cite{SteFr95}, and discussed in detail in \cite{Sterev}. This code was used for the study of different models of rotating neutron stars in \cite{No98} and for the study of the rapidly rotating strange stars (\cite{Ste99}). The software provides the metric potentials for various types of compact rotating general relativistic objects, which can be used to obtain the electron-positron energy deposition rate in the equatorial plane of rapidly rotating neutron and quark stars.

The present paper is organized as follows. In Section 2 we present the basic formalism for the calculation of the electron-positron energy deposition rate from neutrino-antineutrino annihilation. The general relativistic corrections to this process are obtained in Section 3. The equations of state of dense neutron and quark matter used in the present study are presented in Section 4. In Section 5 we obtain the electron-positron energy deposition rates in the equatorial plane of the considered classes of neutron and quark stars. We discuss and conclude our results in Section 6.

\section{Neutrino pair annihilation}

A considerable amount of energy can be released by the neutrino pair annihilation
process in the regions close to the so called neutrino-sphere, with
radius $R_{\nu}$, at which the mean free path of the neutrino
is equal to the radius itself \citep{BeWi85}.

The energy deposition rate per unit volume is given by
\begin{equation}\label{eq:dotq}
\dot{q}(\boldsymbol{r})=\int\int f_{\nu}(\boldsymbol{p}_{\nu},{\boldsymbol{r}})f_{\bar{\nu}}({\boldsymbol{p}}_{_{\bar{\nu}}},{\boldsymbol{r}})\{\sigma|{\boldsymbol{v}}_{\nu}-
{\boldsymbol{v}}_{\bar{\nu}}|\varepsilon_{\nu}\varepsilon_{\bar{\nu}}\}\frac{\varepsilon_{\nu}+
\varepsilon_{\bar{\nu}}}{\varepsilon_{\nu}\varepsilon_{\bar{\nu}}}d^{3}{\boldsymbol{p}}_{\nu}d^{3}
{\boldsymbol{p}}_{\bar{\nu}},
\end{equation}
for any point ${\boldsymbol{r}}>{\boldsymbol{R}}_{\nu}$, where $f_{\nu}$
and $f_{\bar{\nu}}$ are the number densities in the momentum
spaces with the momenta ${\boldsymbol{p}}_{\nu}$ and ${\boldsymbol{p}}_{_{\bar{\nu}}}$, and
${\boldsymbol{v}}_{\nu}$ , $\varepsilon_{\nu}$, ${\boldsymbol{v}}_{\bar{\nu}}$
and $\varepsilon_{\bar{\nu}}$ are the 3-velocities and the energy
of the colliding neutrino-antineutrino pairs, respectively \cite{Go87}.
The cross-section of the collision is denoted by $\sigma $.

By applying the decompositions
${\boldsymbol{p}}_{\nu} =\varepsilon_{\nu }\boldsymbol{\Omega }_{\nu }$
and $d^{3}\boldsymbol{p}_{\nu}=\varepsilon_{\nu }^{2}d\varepsilon_{\nu}d\boldsymbol{\Omega}_{\nu}$,
with the solid angle vector
$\boldsymbol{\Omega}_{\nu }$
pointing
in the direction of ${\boldsymbol{p}}_{\nu}$, and with the assumption that the neutrino-sphere emits particles isotropically, the integral in Eq.~(\ref{eq:dotq}) can be separated into an energy
integral and an angular part. After evaluating the
energy integral for fermions, for a spherically symmetric geometry the
energy deposition rate per unit volume can be represented as
\begin{equation}\label{eq:dotq2}
\dot{q}(r)=\frac{7DG_{F}^{2}\pi^{3}\zeta(5)}{2c^{5}h^{6}}(kT)^{9}\Theta(r)\propto T^{9}(r)\Theta(r),
\end{equation}
where $T$ is the neutrino temperature, $\zeta $ is the Riemann function, $D=1\pm4\sin^{2}\theta_{W}+8\sin^{4}\theta_{W}$, with $\sin^{2}\theta_{W}=0.23$,
and $G_{F}^{2}=5.29\times10^{-44}\;\mathrm{cm}^{2}\;\mathrm{MeV}^{-2}$, respectively, and with the angular part of the energy deposition rate $\Theta(r)$ given by
\begin{equation}\label{eq:THETA}
\Theta(r)=\int\int(1-{\boldsymbol{\Omega}}_{\nu}\boldsymbol{\cdot}\boldsymbol{\Omega}_{\bar{\nu}})^{2}d\boldsymbol{\Omega}_{\nu}d\boldsymbol{\Omega}_{\bar{\nu}}.
\end{equation}
The radial momentum density $\dot{p}$ transported into the $e^+e^-$ plasma from the colliding $\nu\bar{\nu}$ pairs can be written as \citep{SaWi99}
\begin{equation}\label{eq:dotp}
\dot{p}=\frac{\dot{q}}{c}{\Phi_p(r)}{\Theta(r)}\;,
\end{equation}
where
\begin{equation}\label{eq:PHI_p}
\Phi_p(r)=\frac{1}{2}\int\int(1-{\boldsymbol{\Omega}}_{\nu}\boldsymbol{\cdot}\boldsymbol{\Omega}_{\bar{\nu}})^{2}[\boldsymbol{r}\boldsymbol{\cdot}({\boldsymbol{\Omega}}_{\nu}+\boldsymbol{\Omega}_{\bar{\nu}})]d\boldsymbol{\Omega}_{\nu}d\boldsymbol{\Omega}_{\bar{\nu}},
\end{equation}
with the unit vector $\boldsymbol{r}$ normal to the stellar surface.

Since the integrals in Eqs.~(\ref{eq:THETA}) and (\ref{eq:PHI_p}) depend only on the radial coordinate $r$, by virtue of the symmetry, the angular part $\Theta(r)$ and the function $\Phi_p(r)$ can be given by the analytic formulae
\begin{equation}
\Theta(r)=\frac{2\pi^{2}}{3}(1-x)^{4}(x^{2}+4x+5)\label{eq:THETAx}
\end{equation}
and
\begin{equation}
\Phi_p(r)=\frac{\pi^{2}}{6}(1-x)^{4}(8+17x+12x^{2}+3x^3)\;,\label{eq:PHI_px}
\end{equation}
where $x=\cos\theta_{max}$, and $\theta_{max}$ is the maximal angle
between ${\boldsymbol{r}}$ and ${\boldsymbol{p}}_{\nu}$ \citep{Go87, SaWi99}.
In order to determine the quantity $x$, one
need to use the equations of motion of the neutrinos radiated by
the stellar matter, and propagating outside the neutrino-sphere. The
equations of motion in the  Newtonian case, or the geodesic equations in
the general relativistic case fully determine $x$ as a
function of the radial coordinate. Therefore, the properties of the gravitational
potential produce an imprint on the energy deposition rate calculated
in the region close to the neutrino-sphere. Salmonson \& Wilson (1999) extended
the calculations of Goodman et al. (1987) to the case of the Schwarzschild geometry, and compared their results to those of the Newtonian case.

Although rotating configurations of neutron stars break the spherical symmetry
of the space-time, one can still carry out an analysis similar to the static case
if the study is restricted to the investigation of the annihilation of neutrino and antineutrino pairs
propagating in the equatorial plane of the rotating star. By eliminating
the angular dependence from the equations, a formalism similar to the spherically symmetric case can
be used to calculate the energy deposition rate in the equatorial
plane. The obtained result is not equal to the total deposition rate of the high energy electron positron pairs created in the annihilation process. However, this quantity can still be applied in the comparison of the neutrino and antineutrino annihilation energy deposition rate for different models of rotating neutron and quark stars, or in studying the general effects of the rotation of the stellar object on this process.

\section{General relativistic effects on the electron-positron energy deposition rate}

The metric of a stationary and axisymmetric geometry is given in the general form  by
\begin{equation}\label{eq:ds}
ds^{2}=g_{tt}dt^{2}+2g_{t\phi}dtd\phi+g_{rr}dr^{2}+g_{\theta\theta}d\theta^{2}+g_{\phi\phi}d\phi^{2}.
\end{equation}
 For this metric the null-geodesics equations in the equatorial plane $\theta=\pi/2$
are
\begin{eqnarray}
\dot{t} & = & \frac{g_{\phi\phi}E+g_{t\phi}L}{g_{\phi t}^{2}-g_{tt}g_{\phi\phi}},\label{eq:dott}\\
\dot{\phi} & = & -\frac{g_{\phi t}E+g_{tt}L}{g_{t\phi}^{2}-g_{tt}g_{\phi\phi}},\label{dotphi}\\
g_{rr}\dot{r}^{2} & = & \frac{g_{\phi\phi}E^{2}+2g_{t\phi}EL+g_{tt}L^{2}}{g_{t\phi}^{2}-g_{tt}g_{\phi\phi}},\label{eq:dotr}
\end{eqnarray}
where $E$ is the energy, and $L$ is the angular momentum of the particles
propagating along the null-geodesics. Then the parametric equation for $dr/d\phi$
can be written as
\begin{equation}\label{eq:grrdrdphi2}
g_{rr}\left(\frac{dr}{d\phi}\right)^{2}=(g_{t\phi}^{2}-g_{tt}g_{\phi\phi})\frac{g_{\phi\phi}+2g_{t\phi}b+g_{tt}b^{2}}{(g_{t\phi}+g_{tt}b)^{2}},
\end{equation}
where the impact parameter is defined as $b=L/E$.

For the metric given by Eq.~(\ref{eq:ds}), the locally non-rotating frame (LNRF) (in which the wordlines of the freely falling observers are $r=$ constant, $\theta$ = constant and $\phi-\omega t =$ constant, respectively,  with $\omega = -g_{t\phi}/g_{\phi\phi}$) has the basis of one-forms \citep{BPT72}
\begin{eqnarray*}
e_{\mu}^{(t)} & = & \sqrt{\frac{-g_{tt}}{g_{t\phi}^{2}-g_{tt}g_{\phi\phi}}}(-1,0,0,0),\\
e_{\mu}^{(r)} & = & (0,\sqrt{g_{rr}},0,0),\\
e_{\mu}^{(\theta)} & = & (0,0,\sqrt{g_{\theta\theta}},0),\\
e_{\mu}^{(\phi)} & = & \sqrt{g_{\phi\phi}}(g_{t\phi}/g_{\phi\phi},0,0,1).
\end{eqnarray*}

Since the velocity measured in the LNRF is given by $v^{(a)}=e_{\mu}^{(a)}v^{\mu}$, the angle $\theta_{r}$  between
the particle trajectory and the tangent vector to the circular orbit
with the radial coordinate $r$ can be written as
\[
\tan\theta_{r}=\frac{v^{(r)}}{v^{(\phi)}}=\frac{e_{r}^{(r)}v^{r}}{e_{\phi}^{(\phi)}v^{\phi}+e_{t}^{(\phi)}}=\frac{\sqrt{g_{rr}}}{\sqrt{g_{\phi\phi}}[1+g_{t\phi}/(g_{\phi\phi}v^{\phi})]}\frac{dr}{d\phi}.
\]
 From this expression we obtain for $dr/d\phi $ the equation
 \begin{equation}\label{eq:drdphi}
\frac{dr}{d\phi}=\sqrt{\frac{g_{\phi\phi}}{g_{rr}}}\left(1+\frac{g_{t\phi}}{g_{\phi\phi}v^{\phi}}\right)\tan\theta_{r}.
\end{equation}
By inserting Eqs.~(\ref{eq:dott}) and (\ref{dotphi}) into the definition
of $v^{\phi}$, we obtain for $v^{\phi}$ the expression
\[
v^{\phi}=\frac{u^{\phi}}{u^{t}}=\frac{\dot{\phi}}{\dot{t}}=-\frac{g_{t\phi}E+g_{tt}L}{g_{\phi\phi}E+g_{t\phi}L},
\]
 which can be substituted into Eq.~(\ref{eq:drdphi}) to give
 \[
\frac{dr}{d\phi}=\sqrt{\frac{g_{\phi\phi}}{g_{rr}}}\left(1-\frac{g_{t\phi}}{g_{\phi\phi}}\frac{g_{\phi\phi}E+g_{t\phi}L}{g_{t\phi}E+g_{tt}L}\right)\tan\theta_{r}.
\]
Then the derivative $dr/d\phi$ can be eliminated from the parametric
equation (\ref{eq:grrdrdphi2}) and we obtain
\[
g_{\phi\phi}\left(1-\frac{g_{t\phi}}{g_{\phi\phi}}\frac{g_{\phi\phi}+g_{t\phi}b}{g_{t\phi}+g_{tt}b}\right)^{2}\tan^{2}\theta_{r}=(g_{t\phi}^{2}-g_{tt}g_{\phi\phi})\frac{g_{\phi\phi}+2g_{t\phi}b+g_{tt}b^{2}}{(g_{\phi t}+g_{tt}b)^{2}}.
\]
This result gives a second order algebraic equation for $b$,
\begin{equation}\label{eq:gttgppgtp2}
[(g_{tt}g_{\phi\phi}-g_{t\phi}^{2})\sec^{2}\theta_{r}+g_{t\phi}^{2}]b^{2}+2g_{t\phi}g_{\phi\phi}b+g_{\phi\phi}^{2}=0,
\end{equation}
which can be solved to give the impact parameter $b$ as
\begin{equation}\label{eq:bpm}
b_{\pm}=\frac{-g_{\phi\phi}}{g_{t\phi}\pm\sqrt{g_{t\phi}^{2}-g_{tt}g_{\phi\phi}}\sec\theta_{r}}.
\end{equation}

A particular system of coordinates that is used in the study of the general-relativistic rotating configurations is the quasi-isotropic coordinate system $(t,\bar{r},\theta,\phi)$, in which the line element can be represented as \citep{SteFr95,Sterev}
\begin{equation}
ds^2=-e^{\bar{\gamma }+\bar{\rho }}dt^2+e^{2\bar{\alpha } }\left(d\bar{r}^2+\bar{r}^2d\theta ^2\right)+e^{\bar{\gamma }-\bar{\rho }}\bar{r}^2\sin ^2\theta \left(d\phi -\bar{\omega}dt\right)^2,\label{eq:ds2}
\end{equation}
where $\bar{\gamma }$, $\bar{\rho }$, $\bar{\alpha }$ and the angular velocity of the stellar fluid relative to the local inertial frame $\bar{\omega}$ are all functions of the quasi-isotropic radial coordinate $\bar{r}$ and of the polar angle $\theta $.

If for neutron
stars the metric (\ref{eq:ds}) is given in an isotropic coordinate system in the form ({\ref{eq:ds2}}),  then the second order algebraic equation ~(\ref{eq:gttgppgtp2}) for $b$ can be written in terms of the metric functions $\bar{\rho}$ and $\bar{\omega}$ as
\[
\left(\sec^{2}\theta_{r}+e^{-2\rho}\bar{\omega}^{2}\bar{r}^{2}\sin^{2}\theta\right)b^{2}-2e^{-2\bar{\rho}}\bar{\omega}\bar{r}^{2}\sin^{2}\theta b+e^{-2\bar{\rho}}\bar{r}^{2}\sin^{2}\theta=0.
\]
For the impact parameter, corresponding to $\theta_{r}=0$, we obtain \cite{Ca07}
\[
b_{\pm}=\frac{-e^{-\bar{\rho}}\bar{r}\sin\theta}{-e^{-\bar{\rho}}\bar{\omega}\bar{r}\sin\theta\pm\sec\theta_{r}}=\pm\frac{e^{-\bar{\rho}}\bar{r}\sin\theta}{\sec\theta_{r}\pm e^{-\bar{\rho}}\bar{\omega}\bar{r}\sin\theta}.
\]

From the parametric equation Eq.~(\ref{eq:grrdrdphi2}) we obtain the deflection
angle of the particle trajectory for a given $b$ as
\begin{equation}\label{eq:Deltaphi}
\Delta\phi=\int_{r_{em}}^{r_{obs}}\frac{\sqrt{g_{rr}}(g_{t\phi}+g_{tt}b)dr}{\sqrt{(g_{t\phi}^{2}-g_{tt}g_{\phi\phi})(g_{\phi\phi}+2g_{t\phi}b+g_{tt}b^{2})}}.
\end{equation}

In this equation $\Delta\phi$ measures the change in the angle
between the source and the observer, for a photon emitted at the radial coordinate
$r_{em}$, and observed at the radial coordinate $r_{obs}$. This equation
can also be given in the equatorial plane in terms of the metric functions $\bar{\rho}$ and $\bar{\omega}$, respectively, appearing in the line element Eq.~(\ref{eq:ds2}) \cite{Ca05}
\[
\Delta\phi=  -\int_{\bar{r}_{em}}^{\bar{r}_{obs}}e^{\bar{\alpha}-(\bar{\gamma}+\bar{\rho)}/2}\frac{\bar{\omega}(1-\bar{\omega} b)+be^{2\bar{\rho}}/\bar{r}^{2}}{\sqrt{(1-\bar{\omega} b)^{2}-b^{2}e^{2\bar{\rho}}/\bar{r}^{2}}}d\bar{r}
.
\]

In the equatorial plane of black holes the photon radius is defined as the innermost boundary of circular orbits  below which massless particles with $\theta_r=0$ are gravitationally bound \citep{BPT72}. For static black holes the photon radius is $3M$, and for rotating black holes it reduces to $M$, as the spin parameter $a_*=J/M^2$ of the black hole approaches unity. This orbit may exist for ultra-compact stars as well. For static stars with a stellar radius less than $3M$ there is always such a "photon sphere", whereas, depending on the geometry of the space-time, the rotating stars can also have a photon radius at both lower and higher radii \cite{MaMa09}. On the other hand, very massive rotating quark stars in the Color-Flavor-Locked phase can reach masses higher than the equilibrium limit for static stars ($\approx 3M_{\odot}$), of the same order as the stellar mass black holes, and thus they can also have a photon radius \cite{Ko09}.

If the integral in Eq.~(\ref{eq:Deltaphi}) diverges to infinity (or $dr/d\phi$
tends to zero), the null particles are rotating around the central
objects in circular orbits. In this case $\theta_{r}=0$, and the algebraic
equation Eq.~(\ref{eq:gttgppgtp2}) for $b$ reduces to $g_{\phi\phi}+2g_{t\phi}b+g_{tt}b^{2}=0$,
with the solution
\begin{equation}\label{eq:bpm2}
b_{\pm}=\frac{-g_{\phi\phi}}{g_{t\phi}\pm\sqrt{g_{t\phi}^{2}-g_{tt}g_{\phi\phi}}}=\pm\frac{e^{-\bar{\rho}}\bar{r}}{1\pm e^{-\bar{\rho}}\bar{\omega}\bar{r}}.
\end{equation}

For any value of $b$ satisfying Eq.~(\ref{eq:bpm2}), the integral
(\ref{eq:Deltaphi}) is divergent, and the null particles have circular
orbits. In Eq.~(\ref{eq:bpm2}) the impact parameter in the equatorial plane is given as a function of the radial coordinate only.

In the case of the ultracompact static stars with radii $R_e$ less than $3M$ the (local) maximum of the function $b(r)$ is located at $r=3M$, providing the photon radius. If $R_e>3M$, then $b$ is a monotonically decreasing function, without a local maximum. For rotating compact stars, the function $b(r)$ provides the same criterion for the existence of the potential barrier: for massless particles the equatorial orbit where $b(r)$ attains its local maximum defines the innermost boundary of circular orbits. Even if this value is less than the equatorial radius $R_e$, neutrinos are still free to propagate along orbits lying on the photon radius.

By assuming that the mean free path of the neutrinos is equal to or less than the photon radius, Salmonson \& Wilson (1999) identified the photon and the neutrino spheres with each other. Accordingly, we will also consider the orbit at the photon radius as the minimal radius where the annihilation process should still be taken into account, provided it is outside the star. The contribution of the electron-positron pairs formed inside the star to the deposition rate is neglected, because of their complicated interactions with the neutron and quark matter.

Since the impact parameter measured at infinity is constant along
the trajectory of any null particle, the neutrinos propagating from
a point at the photon sphere radius $R$ (or the stellar surface $R_e$) with the angle $\theta_{R}=\tan^{-1}v^{(r)}(R)/v^{(\phi)}(R)$,
will reach another point with radial coordinate $r$ with the angle
\begin{equation}\label{angle}
\cos\theta_{r}=\frac{g_{\phi\phi}(R)\sqrt{g_{t\phi}^{2}(r)-g_{tt}(r)g_{\phi\phi}(r)}}{g_{\phi\phi}(r)\left[g_{t\phi}(R)+\sqrt{g_{t\phi}^{2}(R)-g_{tt}(R)g_{\phi\phi}(R)}\sec\theta_{R}\right]-g_{\phi\phi}(R)g_{t\phi}(r)}.
\end{equation}
Eq.~(\ref{angle}) allows to express $x$ in the analytic expression of the angular part $\Theta(r)$ of the energy, given by Eq.~(\ref{eq:THETAx}), as
\begin{equation}\label{eq:x2}
x^{2}(r)=1-\frac{g_{\phi\phi}^{2}(R)\left[g_{t\phi}^{2}(r)-g_{tt}(r)g_{\phi\phi}(r)\right]}{\left\{ g_{\phi\phi}(r)\left[g_{t\phi}(R)+\sqrt{g_{t\phi}^{2}(R)-g_{tt}(R)g_{\phi\phi}(R)}\right]-g_{\phi\phi}(R)g_{t\phi}(r)\right\} ^{2}}.
\end{equation}

For the line element given by Eq.~(\ref{eq:ds2}), Eq.~(\ref{eq:x2}) has the form
\begin{equation}
x^{2}(\bar{r})=1-\left\{\frac{\bar{R}}{\bar{r}}\frac{e^{\gamma(\bar{r})-\gamma(\bar{R})}}{1+[\omega(\bar{r})-\omega(\bar{R})]\bar{R}e^{\gamma(\bar{R})}}\right\}^2\:.
\end{equation}

Thus the function $\Theta (r)$ can be represented in the general form
\begin{eqnarray}
\Theta(r) & = & \frac{2\pi^{2}}{3}\left[1-\sqrt{1-\frac{g_{\phi\phi}^{2}(R)\left[g_{t\phi}^{2}(r)-g_{tt}(r)g_{\phi\phi}(r)\right]}{\left\{ g_{\phi\phi}(r)\left[g_{t\phi}(R)+\sqrt{g_{t\phi}^{2}(R)-g_{tt}(R)g_{\phi\phi}(R)}\right]-g_{\phi\phi}(R)g_{t\phi}(r)\right\} ^{2}}}\right]^{4}\nonumber\\
 &  & \times\left[-\frac{g_{\phi\phi}^{2}(R)\left[g_{t\phi}^{2}(r)-g_{tt}(r)g_{\phi\phi}(r)\right]}{\left\{ g_{\phi\phi}(r)\left[g_{t\phi}(R)+\sqrt{g_{t\phi}^{2}(R)-g_{tt}(R)g_{\phi\phi}(R)}\right]-g_{\phi\phi}(R)g_{t\phi}(r)\right\} ^{2}}\right.\nonumber\\
 &  & \left.+4\sqrt{1-\frac{g_{\phi\phi}^{2}(R)\left[g_{t\phi}^{2}(r)-g_{tt}(r)g_{\phi\phi}(r)\right]}{\left\{ g_{\phi\phi}(r)\left[g_{t\phi}(R)+\sqrt{g_{t\phi}^{2}(R)-g_{tt}(R)g_{\phi\phi}(R)}\right]-g_{\phi\phi}(R)g_{t\phi}(r)\right\} ^{2}}}+6\right].
 \end{eqnarray}
 In quasi-isotropic coordinates we have
\begin{eqnarray}
\Theta(\bar{r}) & = & \frac{2\pi^{2}}{3}\left[1-\sqrt{1-\left\{ \frac{\bar{R}}{\bar{r}}\frac{e^{\gamma(\bar{r})-\gamma(\bar{R})}}{1+\left[\omega(\bar{r})-\omega(\bar{R}))\right]\bar{R}
e^{\gamma(\bar{R})}}\right\} ^{2}}\right]^{4}\nonumber\\
 &  & \times\left[\left\{ \frac{\bar{R}}{\bar{r}}\frac{e^{\gamma(\bar{r})-\gamma(\bar{R})}}{1+\left[\omega(\bar{r})-\omega(\bar{R})\right]\bar{R}
 e^{\gamma(\bar{R})}}\right\} ^{2}+4\sqrt{1-\left\{ \frac{\bar{R}}{\bar{r}}\frac{e^{\gamma(\bar{r})-\gamma(\bar{R})}}{1+\left[\omega(\bar{r})-\omega(\bar{R}))\right]
 \bar{R}e^{\gamma(\bar{R})}}\right\} ^{2}}+6\right].\nonumber\\
 \end{eqnarray}

The neutrino temperature at the radius $r$ can be expressed in terms of the temperature of the neutrino stream at
the neutrino-sphere radius $R$, by taking into account the gravitational redshift. The redshift formula for $T$ is the same as for the photon energy,
\begin{equation}
T(r)=\left\{ \frac{g_{\phi\phi}(r)\left[g_{t\phi}^{2}(R)-g_{tt}(R)g_{\phi\phi}(R)\right]}{g_{\phi\phi}(R)\left[g_{t\phi}^{2}(r)-g_{tt}(r)g_{\phi\phi}(r)\right]}\right\} ^{1/2}T(R)\:.\label{eq:Tr}\end{equation}
For the observed luminosity $L_{\infty}$ of the neutrino annihilation the
redshift relation is given by \begin{equation}
L_{\infty}=\frac{g_{\phi\phi}(r\rightarrow\infty)\left[g_{t\phi}^{2}(R)-g_{tt}(R)g_{\phi\phi}(R)\right]}{g_{\phi\phi}(R)\left[g_{t\phi}^{2}(r\rightarrow\infty)-g_{tt}(r\rightarrow\infty)g_{\phi\phi}(r\rightarrow\infty)\right]}L(R)=\left[\frac{g^2_{t\phi}(R)}{g_{\phi\phi}(R)}-g_{tt}(R)\right]L(R)\:,\label{eq:Lr}\end{equation}
since for isolated gravitating systems, such as rotating stars, the spacetime is asymptotically flat. Here the neutrino luminosity at the neutrino-sphere is
\begin{equation}
L(R)=L_{\nu}+L_{\bar{\nu}}=(4\pi R^{2})\frac{7}{16}acT^{4}(R)\label{eq:LR}\end{equation}
where $a$ is the radiation constant. In this formula the curvature radius $R$ is used to obtain the total area of the spherical surface through which the neutrino radiation is emitted. If we insert Eq.~(\ref{eq:x2}), describing the path-bending of the neutrinos,  and the redshift formulae Eqs.~ (\ref{eq:Tr})-(\ref{eq:LR})
into the decomposed expression Eq.~(\ref{eq:dotq2}) of $\dot{q}$, we
can calculate the effects of the gravitational potential on the deposition
rate in the equatorial plane:
\begin{equation}\label{eq:dotq3}
\dot{q}(r)\propto L^{9/4}_{\infty}\Theta(r)
\left\{\frac{g_{\phi\phi}(r)\sqrt{g_{t\phi}^{2}(R)-g_{tt}(R)g_{\phi\phi}(R)}}{\sqrt{g_{\phi\phi}(R)}\left[g_{t\phi}^{2}(r)-g_{tt}(r)g_{\phi\phi}(r)\right]}\right\} ^{9/2}R^{-9/4}.
\end{equation}

The proportionality factor, omitted from Eq.~(\ref{eq:dotq3}), is the same as the one in the Newtonian case, which will be used in the following to normalize the deposition rate for the general relativistic case.
Eq.~(\ref{eq:dotq3}) describes the energy deposition rate in $e^+e^-$ pairs from the neutrino-antineutrino annihilation process at radius $r$ in the equatorial plane above the neutron or quark star photon sphere radius $R$, and with the neutrino luminosity observed at infinity $L_{\infty}$. This relation can be also given in the quasi-isotropic coordinate system in terms of the metric functions of the line element Eq.~(\ref{eq:ds2}),
\begin{equation}
\dot{q}(\bar{r})\propto L^{9/4}_{\infty}\Theta(\bar{r}) e^{9[\gamma(\bar{R})+\rho(\bar{R})]/4-9[\gamma(\bar{r})+\rho(\bar{r})]/2}
R^{-9/4}(\bar{R}).
\end{equation}

\section{Equations of state and stellar models}\label{eos}

In order to obtain a consistent and realistic physical description of the rotating general relativistic neutron and quark stars, as a first step we have to adopt the equations of state for the dense neutron and quark matter, respectively. In the present study we consider the following equations of state for neutron and quark matter:

1) Akmal-Pandharipande-Ravenhall  (APR) EOS \citep{Ak98}. EOS APR has been obtained by using the variational chain summation methods and the  Argonne $v_{18}$ two-nucleon interaction. Boost corrections to the two-nucleon interaction, which give the leading relativistic effect of order $(v/c)^2$, as well as three-nucleon interactions, are also included in the nuclear Hamiltonian.  The density range is from $2\times 10^{14}$ g/cm$^3$ to $2.6\times 10^{15}$ g/cm$^3$. The maximum mass limit in the static case for this EOS is $2.20 M_{\odot}$.  We join this equation of state to the composite BBP ($\epsilon /c^2>4.3\times10^{11}$g/cm$^3$) \citep{Ba71a} - BPS ($10^4$ g/cm$^3$ $<4.3\times 10^{11}$g/cm$^3$) \citep{Ba71b} - FMT ($\epsilon/c^2<10^4$ g/cm$^3$) \citep{Fe49} equations of state, respectively.

2) Douchin-Haensel (DH) EOS \citep{DoHa01}. EOS DH is an equation of state of the neutron star matter, describing both the neutron star crust and the liquid core. It is based on the effective nuclear interaction SLy of the Skyrme type, which is particularly suitable for the application to the calculation of the properties of very neutron rich matter. The structure of the crust, and its EOS, is calculated in the zero temperature approximation, and under the assumption of the ground state composition. The EOS of the liquid core is calculated assuming (minimal) $npe\mu $ composition. The density range is from $3.49\times 10^{11}$ g/cm$^3$ to $4.04\times 10^{15}$ g/cm$^3$. The minimum and maximum masses of the static neutron stars for this EOS are $0.094M_{\odot}$ and $2.05 M_{\odot}$, respectively.

3) 	Shen-Toki-Oyamatsu-Sumiyoshi (STOS) EOS \citep{Shen}. The STOS equation of state of nuclear matter is obtained by using the relativistic mean field theory with nonlinear $\sigma $ and $\omega $ terms in a wide density and temperature range, with various proton fractions. The EOS was specifically designed for the use of supernova simulation and for the neutron star calculations. The  Thomas-Fermi approximation is used to describe inhomogeneous matter, where heavy nuclei are formed together with free nucleon gas. We consider the STOS EOS for several temperatures, namely $T=0$, $T=0.5$ and $T=1.0$ MeV, respectively. The temperature is mentioned for each STOS equation of state, so that, for example, STOS 0 represents the STOS EOS for $T=0$. For the proton fraction we chose  the value $Y_p=10^{-2}$ in order to avoid the negative pressure regime for low baryon mass densities.

4) Relativistic Mean Field (RMF) equations of state with isovector scalar mean field corresponding to the $\delta $-meson- RMF soft and RMF stiff EOS \citep{Kubis}.  While the $\delta $-meson mean field vanishes in symmetric nuclear matter, it can influence properties of asymmetric nuclear matter in neutron stars. The Relativistic mean field contribution due to the $\delta $-field to the nuclear symmetry energy is negative. The energy per particle of neutron matter is then larger at high densities than the one with no $\delta $-field included. Also, the proton fraction of $\beta $-stable matter increases. Splitting of proton and neutron effective masses due to the $\delta $-field can affect transport properties of neutron star matter. The equations of state can be parameterized by the coupling parameters $C_{\sigma }^2=g_{\sigma }^2/m_{\sigma }^2$, $C_{\omega }^2=g_{\omega }^2/m_{\omega }^2$, $\bar{b}=b/g_{\sigma}^3$ and $\bar{c}=c/g_{\sigma}^4$, where $m_{\sigma }$ and $m_{\omega }$ are the masses of the respective mesons, and $b$ and $c$ are the coefficients in the potential energy $U\left(\sigma \right)$ of the $\sigma $-field. The soft RMF EOS is parameterized by $C_{\sigma }^2=1.582$ fm$^2$, $C_{\omega }^2=1.019$ fm$^2$, $\bar{b}=-0.7188$ and $\bar{c}=6.563$, while the stiff RMF EOS is parameterized by $C_{\sigma }^2=11.25$ fm$^2$, $C_{\omega }^2=6.483$ fm$^2$, $\bar{b}=0.003825$ and $\bar{c}=3.5\times 10^{-6}$, respectively.

5) Baldo-Bombaci-Burgio (BBB) EOS \citep{baldo}. The BBB EOS is an EOS for asymmetric nuclear matter, derived from the Brueckner-Bethe-Goldstone many-body theory with explicit three-body forces. Two EOS's are obtained, one corresponding to the Argonne AV14 (BBBAV14), and the other to the Paris two-body nuclear force (BBBParis), implemented by the Urbana model for the three-body force. The maximum static mass configurations are $M_{max} = 1.8 M_{\odot}$ and  $M_{max} = 1.94 M_{\odot}$ when the AV14 and Paris interactions are used, respectively.  The onset of direct Urca processes occurs at densities $n\geq 0.65$ fm$^{-3}$ for the AV14 potential and $n\geq 0.54$ fm$^{-3}$ for the Paris potential. The comparison with other microscopic models for the EOS shows noticeable differences. The density range is from $1.35\times 10^{14}$ g/cm$^3$ to $3.507\times 10^{15}$ g/cm$^3$.

6) Bag model equation of state for quark matter (Q) EOS \citep{It70, Bo71, Wi84, Ch98}. For the description of the quark matter we adopt first a simple phenomenological description, based on the MIT bag model equation of state, in which the pressure $p$ is related to the energy density $\rho $ by
\begin{equation}
p=\frac{1}{3}\left(\rho-4B\right)c^2,
\end{equation}
where $B$ is the difference between the energy density of the perturbative and
non-perturbative QCD vacuum (the bag constant), with the value $4B=4.2\times 10^{14}$ g/cm$^3$.

7) It is generally agreed today that the color-flavor-locked (CFL) state is
likely to be the ground state of matter, at least for asymptotic densities,
and even if the quark masses are unequal \citep{cfl1,cfl2, HoLu04, cfl3}. Moreover,
the equal number of flavors is enforced by symmetry, and electrons are
absent, since the mixture is automatically neutral. By assuming that the
mass $m_{s}$ of the $s$ quark is not large as compared to the chemical
potential $\mu $, the thermodynamical potential of the quark matter in CFL
phase can be approximated as \citep{LuHo02}
\begin{equation}
\Omega _{CFL}=-\frac{3\mu ^{4}}{4\pi ^{2}}+\frac{3m_{s}^{2}}{4\pi ^{2}}-%
\frac{1-12\ln \left( m_{s}/2\mu \right) }{32\pi ^{2}}m_{s}^{4}-\frac{3}{\pi
^{2}}\Delta ^{2}\mu ^{2}+B,
\end{equation}%
where $\Delta $ is the gap energy. With the use of this expression the
pressure $P$ of the quark matter in the CFL phase can be obtained as an
explicit function of the energy density $\varepsilon $ in the form \citep%
{LuHo02}
\begin{equation}\label{pres}
P=\frac{1}{3}\left( \varepsilon -4B\right) +\frac{2\Delta ^{2}\delta ^{2}}{\pi
^{2}}-\frac{m_{s}^{2}\delta ^{2}}{2\pi ^{2}},
\end{equation}
where
\begin{equation}
\delta ^{2}=-\alpha +\sqrt{\alpha ^{2}+\frac{4}{9}\pi ^{2}\left( \varepsilon
-B\right) },
\end{equation}%
and $\alpha =-m_{s}^{2}/6+2\Delta ^{2}/3$. In the following the value of the gap energy $\Delta $ considered in each case will be also mentioned for the CFL equation of state, so that, for example, CFL200 represents the CFL EOS with $\Delta =200$. For the bag constant $B$ we adopt the value $4B=4.2\times 10^{14}$ g/cm$^3$, while for the mass of the strange quark we take the value $m_s=150$ MeV.


The pressure-density relation is presented for the considered equations of state in Fig.~{\ref{fig1}.

\begin{figure}[tbp]
\centering
\includegraphics[width=8.15cm]{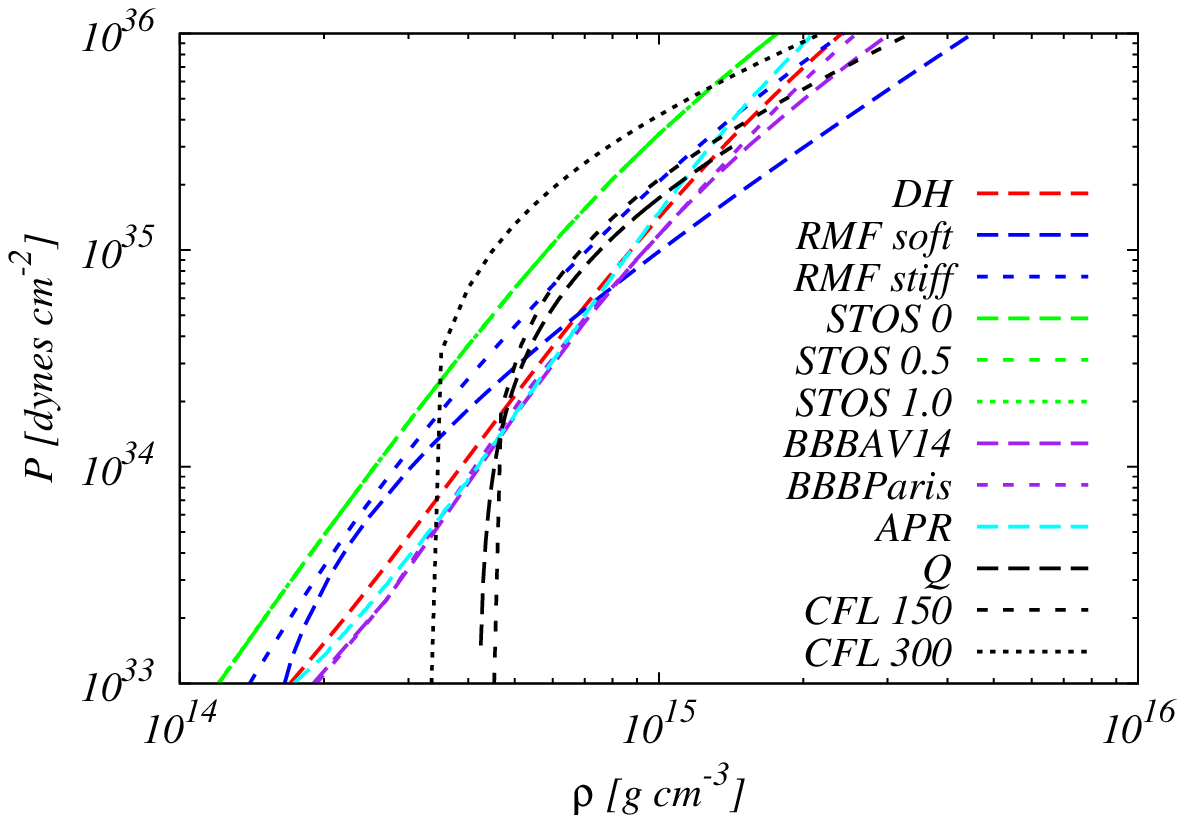}
\caption{Pressure as a function of density (in a logarithmic scale) for the equations of state  DH, RMF soft, RMF stiff, STOS 0. STOS 0.5, STOS 1, BBBAV14, BBBParis, APR, Q,  CFL150, and CFL300, respectively.}
\label{fig1}
\end{figure}

 To calculate the equilibrium configurations of the rotating neutron and quark stars with the EOS's presented here  we use the RNS code, as introduced in Stergioulas \& Friedman (1995), and discussed in details in Stergioulas (2003). This code was used for the study of different models of  rotating neutron stars \citep{No98}, and for the study of the rapidly rotating strange stars \citep{Ste99}. The RNS code produces the metric functions in a quasi-spheroidal coordinate system, as functions of the parameter $s=\bar{r}/\left(\bar{r}+\bar{r}_e\right)$, where $\bar{r}_e$ is the equatorial radius of the star, which we have converted into Schwarzschild-type coordinates $r$ according to the equation $r=\bar{r}\exp\left[\left(\bar{\gamma }-\bar{\rho }\right)/2\right]$.

\section{Electron-positron energy deposition rate in the equatorial plane of rapidly rotating neutron and quark stars}

To demonstrate the existence and the location of the photon radius for neutron and quark stars in the static and the rotating cases, respectively, in Fig.~\ref{fig1b} we present $b(r)/M$ as a function of the  Schwarzschild coordinate radius (normalized to the mass of the star).
For the static neutron star (modeled, for example, by the STOS 0.5 EOS), the exterior geometry is always the Schwarzschild one, and the curve $b(r)$ connects to the curve of the static black hole at a radius somewhat lower than $3M$. Then, the curve has a local minimum at $3M$, similarly to the case of the function $b(r)$, corresponding to the Schwarzschild case \citep{LaLi}. Therefore, the star has a photon radius at $3M$. This is not the case for the static quark star (with an EOS of the CFL 300 type), which is not so compact, and its curve $b(r)$ reaches the Schwarzschild black hole curve at a somewhat higher radius than $3M$. The radial dependence of $b(r)$ for rotating stars has also these two features: the metric potentials of the neutron star with RMF stiff type EOS and the Q and CFL type quark stars give local minima for $b(r)$, whereas the neutron star with STOS 0.5 type EOS is not compact enough to have a photon radius (see Table~\ref{table3}).
We note that the photon radii of the rotating stellar objects are located at higher radii than $3M$, which is opposite to the behavior of the rotating black holes, where the photon radii approach $M$ as the black hole spins up.

\begin{figure}
\centering
\includegraphics[width=8.15cm]{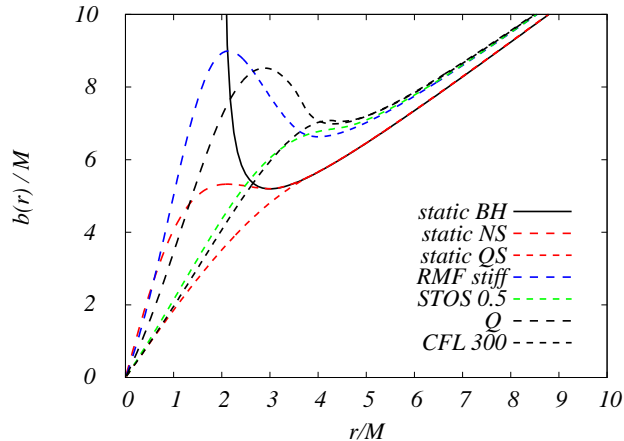}
\caption{The impact parameter $b/M$ as a function of the Schwarzschild coordinate $r/M$ for static black holes, and static and rotating neutron and quark stars, respectively.}
\label{fig1b}
\end{figure}

The deposition rate of the total energy of the $e^{+}e^{-}$ pairs generated
in the neutrino annihilation is usually characterized by the integral
of $\dot{q}$ over the spatial proper volume of the neutrino stream.
The integral of $\dot{q}$ in the radial direction, measuring the total
amount of energy converted form neutrinos to electron-positron pairs
at all radii $R$, is then given by
\begin{eqnarray}\label{eq:dotQ}
\dot{Q}(R) & = & 2\int_{0}^{2\pi}\int_{0}^{\pi/2}\int_{R}^{\infty}\dot{q}(r,R,\theta)\sqrt{g_{rr}g_{\theta\theta}g_{\phi\phi}}drd\theta d\phi.
\end{eqnarray}

For a spherically symmetric geometry the integration over the angular
coordinates $\phi$ and $\theta$ is a straightforward computation.
This is not the case in the axially symmetric case, where the $\theta$-dependence
of the metric as well as the nature of the null trajectories is much more complicated.
In order to avoid this problem, we restrict the study of the electron-positron energy
deposition rate to neutrino pairs moving in the equatorial plane only. Instead
of the quantity $\dot{Q}$ defined by Eq.~(\ref{eq:dotQ}), we consider
its derivative with respect to $\theta $ in the equatorial plane,
\begin{eqnarray}\label{eq:ddotQdth}
\left.\frac{d\dot{Q}}{d\theta}\right|_{\theta=\pi/2} & = & \left.2\int_{0}^{2\pi}\int_{R}^{\infty}\dot{q}(r,R,\theta)\sqrt{g_{rr}g_{\theta\theta}g_{\phi\phi}}drd\phi\right|_{\theta=\pi/2}\nonumber \\
 & = & \left.4\pi\int_{R}^{\infty}\dot{q}(r,R,\theta)\sqrt{g_{rr}g_{\theta\theta}g_{\phi\phi}}dr\right|_{\theta=\pi/2}.
 \end{eqnarray}

$\dot{Q}$, and therefore its $\theta$-derivative,
is neither an observable, nor a Lorentz-invariant quantity \citet{SaWi99}, but $\dot{Q}$
can be used to measure the total amount of {\it local} energy deposited
via the neutrino pair annihilation outside the neutrino-sphere.
The derivative $d\dot{Q}/d\theta$, evaluated at $\theta=\pi/2$ allows
one to describe the energy deposition rate only in the equatorial plane,
but its value normalized to the Newtonian case
can be applied to compare the energy deposition rates in different
gravitational potentials. Since the quantity given by Eq.~(\ref{eq:ddotQdth})
has a simple form in the Newtonian case, the ratio of the general
relativistic case to the Newtonian model is given by
\begin{equation}\label{eq:ddotQdthddotQNdth}
\left.\frac{d\dot{Q}/d\theta}{d\dot{Q}_{N}/d\theta}\right|_{\theta=\pi/2}=\frac{\left.\int_{R}^{\infty}\dot{q}(r,R,\theta)\sqrt{g_{rr}(r,\theta)g_{\theta\theta}(r,\theta)g_{\phi\phi}(r,\theta)}dr\right|_{\theta=\pi/2}}{\int_{R}^{\infty}\dot{q}_{N}(r,R)r^{2}dr}.
\end{equation}

Here $\dot{q}(r,R,\theta)$ is the deposition rate calculated by taking into account the
general relativistic effects, whereas $\dot{q}_{N}(r,R)$ is the deposition rate for the simple
Newtonian case, without taking into account the bending of the neutrino path and the redshift in
the neutrino temperature. We will use the ratio given by Eq.~(\ref{eq:ddotQdthddotQNdth})
to compare the electron-positron energy deposition rates in the space-times
of neutron and quark stars, with different equations of state.

\begin{figure}
\includegraphics[width=8.15cm]{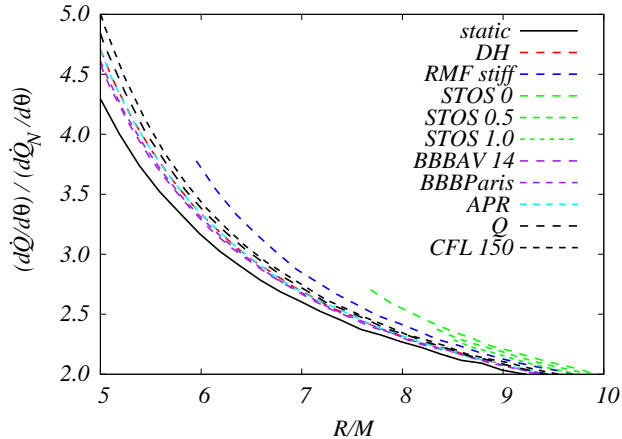}
\centering
\caption{The electron-positron energy deposition rate $\left.\left(d\dot{Q}/d\theta\right)/\left(d\dot{Q}_{N}/d\theta\right)\right|_{\theta=\pi/2}$ in the equatorial plane around rotating neutron and quark stars, with the same total mass $M=1.8M_{\odot}$, and same rotational velocity $\Omega =5\times10^3\;{\rm s}^{-1}$.}
\label{fig2}
\end{figure}

In Fig.~\ref{fig2} we present the ratio given by Eq.~(\ref{eq:ddotQdthddotQNdth}) as a function
of the neutrino-sphere radius, measured in curvature coordinates, for
each neutron and quark star model previously described. The total
mass $M$ of the central objects is set to $1.8M_{\odot}$, and
the rotational frequency $\Omega $ of the stars is about $5\times10^{3}\;\mathrm{s}^{-1}$.
For this configuration the physical parameters of the stars are
shown in Table~\ref{table1}.

\begin{table*}
\centering
\begin{tabular}{|l|l|l|l|l|l|l|l|l|l|l|}
\hline
EOS & DH  &  RMF stiff & STOS 0 & STOS 0.5 & STOS 1 & BBBAV14 & BBBParis & APR & Q & CFL150 \\
\hline
$\rho_c\;[10^{15}{\rm g}/{\rm cm}^{3}]$ & 1.29 & 0.57 & 0.369 & 0.383 & 0.40 & 2.15 & 1.70 & 1.225 &  0.931 & 0.71  \\
\hline
$M\;[M_{\odot}]$ & 1.81  & 1.80 & 1.85 & 1.80 & 1.79 & 1.80 & 1.80 & 1.80 & 1.79 & 1.80\\
\hline
$M_0\; [M_{\odot}]$ & 2.05  & 2.00 & 2.01 & 1.95 & 1.93 & 2.08 & 2.07 & 2.11 &  2.09 & 2.09\\
\hline
$R_e [{\rm km}]$ & 12.01  & 15.79 & 21.03 & 22.84 & 22.72 & 10.57 & 10.98 & 10.99 &  11.79 & 12.36 \\
\hline
$\Omega [10^3{\rm s}^{-1}]$ & 4.99  & 5.00 & 4.90 & 4.71 & 4.45 & 5.01 & 5.00 & 5.00 &  4.79 & 5.00\\
\hline
$\Omega_p [10^3{\rm s}^{-1}]$ & 11.16 &  7.97 & 5.37 & 4.56 & 4.54 & 13.96 & 13.19 & 13.23 &  11.97 & 11.28\\
\hline
$T/W [10^{-2}]$ & 3.43 & 8.93 & 14.91 & 12.25 & 9.83 & 2.35 & 2.66 & 3.41 & 4.45 & 5.64 \\
\hline
$cJ/GM_{\odot}^2$ & 1.15  & 2.00 & 2.95 & 2.49 & 2.15 & 0.95 & 1.00 & 1.12 & 1.28 & 1.50 \\
\hline
$I [10^{45}{\rm g}\;{\rm cm}^2]$ & 2.03  & 3.52 & 5.30 & 4.65 & 4.25 & 1.66 & 1.76 & 1.98 & 2.35 & 2.63 \\
\hline
$\Phi_2 [10^{43}{\rm g}\;{\rm cm}^2]$ & 8.54 & 43.82 & 106.53 & 80.44 & 61.22 & 4.42 & 5.43 & 7.87 & 1.30 & 1.91\\
\hline
$h_+ [{\rm km}]$ & 6.85  & 0.00 & 0.00 & 0.00 & 0.00 & 3.19 & 2.69 & 0.00 &  0.00 & 0.00 \\
\hline
$h_- [{\rm km}]$ &7.48  & -3.40 & 7.91 & 3.87 & 2.27 & 8.14 & 7.90 & -2.11 &  0.00 & 0.00 \\
\hline
$\omega _c/\Omega [10^{-1}]$ & 5.85 & 4.52 & 4.10 & 4.07 & 4.08 & 6.67 & 6.34 & 5.86 &  5.27 & 5.00 \\
\hline
$r_e  [{\rm km}]$ & 9.10 & 12.85 & 18.00 & 19.97 & 19.91 & 7.64 & 8.06 & 8.05 &  8.86 & 9.40\\
\hline
$r_p/r_e$ & 0.88 & 0.72 & 0.54 & 0.53 & 0.58 & 0.92 & 0.91 & 0.90 & 0.87 & 0.84 \\
\hline
\end{tabular}
\caption{Physical parameters of the compact stars with total mass $M\approx 1.8 M_{\odot}$ and rotational frequency $\Omega\approx 5\times 10^3 \;{\rm s}^{-1}$. Here $\rho _c$ is the central density, $M$ is the gravitational mass, $M_0$ is the rest mass, $R_e$ is the circumferential radius at
the equator, $\Omega $ is the angular velocity, $\Omega _p$ is the angular velocity of a particle in circular orbit at the equator, $T/W$ is the rotational-gravitational energy ratio, $cJ/GM_{\odot}^2$ is the angular momentum, $I$ is the moment of inertia, $\Phi_2$ gives the mass quadrupole moment $M_2$ so that $M_2=c^4\Phi_2/(G^2M^3M^3_{\odot})$, $h_+$ is the height from the surface of the last stable co-rotating circular orbit in the equatorial plane, $h_{-}$ is the height from surface of the last stable counter-rotating circular orbit in the equatorial plane, $\omega _c/\Omega$ is the ratio of the central value of the potential $\omega $ to $\Omega $, $r_e$ is the coordinate equatorial radius, and $r_p/r_e$ is the axes ratio (polar to equatorial), respectively.}
\label{table1}
\end{table*}

From the analysis of the compactness of these stars, that is, of the ratio of the total mass to the equatorial
radius, we see that the $1.8M_{\odot}$ mass stars are not compact enough to have a photon
or a neutrino-sphere. In other words, the impact parameter $b$ given by Eq.~(\ref{eq:bpm2}) is a monotonic function of the radial coordinate, without any minimum. For comparison we also plotted the static case
with the same total mass, which is essentially the same for any type
of the equation of state describing the properties of the dense star matter, since the exterior metric of the static stars is described by the Schwarzschild metric.
The energy deposition rate for this case coincides with the $\dot{Q}/\dot{Q}_{N}$ versus $r/M$ plot
given by \citep{SaWi99}, since the multiplicative factor coming
from the integration of $\dot{q}\sqrt{g_{rr}g_{\theta\theta}g_{\phi\phi}}$
over $\theta$ is unity for the Schwarzschild space-time. From Fig.~\ref{fig2} it
can be immediately seen how the energy deposition rate of the electron-positron
pairs is enhanced by the rotation of the central object. It is also
clear that the measure of enhancement depends on the type of the equation
of state used to describe the dense neutron and quark stars. In the rotating case, the increase in the ratio
of the energy deposition rates, given by Eq.~(\ref{eq:ddotQdthddotQNdth}), is the smallest, as compared to the static case, for the BBBAV14 and BBBParis type EOS's. The neutron stars with DH and APR type EOS's produce roughly the same ratios, which are somewhat greater than the BBB type neutron star pair production rates. For the
quark stars, described by the Q and the CFL type EOS's, we obtain even
higher deposition rates. Although none of these types of central
objects possesses neutrino-spheres, they are still compact enough to
have equatorial radii less than $5M$. The equatorial radii of the
neutron stars with the RMF and STOS type EOS's are much larger than
those of the previous group: $6M$ for the RMF type EOS, and $8M$
or even more for the STOS type stars, depending on the temperature
of these stellar configurations. This means that although they give the
highest deposition rates, with the STOS $T=1$ Mev type star having the maximal
value, the energy released by the neutrino pair annihilation outside
the star in the equatorial plane is smaller than in the case of
the first group. This result might be not true in the regions close to the poles
of the star, if the axis ratios $\bar{r}_{p}/\bar{r}_{e}$
would be considerably smaller for the second group as compared to the axis ratios of
the first group. In the case of the STOS type EOS $\bar{r}_{p}/\bar{r}_{e}$
is between 0.5 and 0.6, which is indeed much smaller than 0.85-0.9,
the average values of the axis ratios for the first group. This means
the difference between the polar radii $\bar{r}_{p}$ of the
fist group, and the STOS type neutron star, is smaller than the difference
between their equatorial radii, and the lower boundary of the integral
of $\dot{q}$ over the radial coordinate are closer to each other.
In this case, the integrated deposition rate can be higher for
the neutron star with STOS type EOS than the one with the other types
of equations of state for neutron stars and quark stars.

However, in this framework one should be very
cautious with any statement on the physical processes located far
from the equatorial plane, since the neutrinos reaching
the region close to the pole of the rotating stars have impact parameters
rather different from the neutrinos moving in the equatorial
plane, and the formalism applied for the latter cannot be extended
straightforwardly to the motion outside the equatorial plane.

\begin{table*}
\centering
\begin{tabular}{|l|l|l|l|l|}
\hline
EOS &  RMF stiff &  STOS 1.0 & Q & CFL 300\\
\hline
$\rho_c\;[10^{15}{\rm g}/{\rm cm}^{3}]$ & 2.10 &  0.640000 & 1.60000 & 0.394500\\
\hline
$M\;[M_{\odot}]$ &  2.79986 & 2.8098 & 2.80017 & 2.80677\\
\hline
$M_0\; [M_{\odot}]$ &  3.28120 & 3.19101 &  3.34023 & 3.28947\\
\hline
$R_e [{\rm km}]$ & 14.8969 & 19.6359 & 15.4159 & 18.0389\\
\hline
$\Omega [10^3{\rm s}^{-1}]$ & 9.78914 & 5.90127 & 10.0016 & 5.96099\\
\hline
$\Omega_p [10^3{\rm s}^{-1}]$ & 10.4349 & 6.99013 & 10.4087 & 8.35765\\
\hline
$T/W [10^{-1}]$  & 1.30916 & 1.09688 & 2.00292 & 1.50674\\
\hline
$cJ/GM_{\odot}^2$ & 5.49201 & 5.05616 & 6.64450 & 6.02253\\
\hline
$I [10^{45}{\rm g}\;{\rm cm}^2]$ & 4.93060 & 7.52988 &  5.83855 & 8.87919\\
\hline
$\Phi_2 [10^{44}{\rm g}\;{\rm cm}^2]$ &  5.68507 & 8.48428  & 9.63523 & 13.4904\\
\hline
$h_+ [{\rm km}]$ &  1.93604 & 0.306693 & - & 0.00000\\
\hline
$h_- [{\rm km}]$ &  19.6015 & 15.1710 & 21.4442 & -\\
\hline
$\omega _c/\Omega [10^{-1}]$ &  8.23642 & 6.2819 & 7.57682 &  5.75394\\
\hline
$r_e  [{\rm km}]$ &  9.99343 & 14.9859 & 10.2897 & 13.1041\\
\hline
$r_p/r_e$ &  0.590234 & 0.640000 & 0.502344 & 0.620000\\
\hline
\end{tabular}
\caption{Physical parameters of the compact stars with total mass $M\approx 2.8 M_{\odot}$ and rotational frequency $\Omega\approx 6\times 10^3 \;{\rm s}^{-1}$ and $10^4 \;{\rm s}^{-1}$.}
\label{table2}
\end{table*}

Next, we consider the electron-positron energy deposition rate for a more massive group of stars,
with the total mass $M$ set to $2.8M_{\odot}$. Although this value of the mass is smaller than the
theoretical stability mass limit of 3 Solar masses for ultra-compact
objects, not all types of the EOS's considered here do have configurations
of such high masses, even in the rapidly rotating case. We have
obtained solutions to the field equations for this mass regime only
for the quark stars, for the RMF, and for the STOS type neutron star EOS's, respectively.
For the other equations of state we did not even find a compatible rotation frequency
in the high mass regime. For the stable massive configurations the angular velocity varied
up to $6\times10^{3}\;\mathrm{s}^{-1}$ for the STOS type neutron star
models and the CFL type quark stars, whereas $\Omega$ reached
values of around $9\times10^{3}-10^{4}\;\mathrm{s}^{-1}$ for the neutron
stars with the RMF type EOS, and for Q type quark stars. The models we considered for the study of the electron-positron energy deposition rate have a rotational
frequency of $6\times10^{3}\;\mathrm{s}^{-1}$, for the first two models,
and a frequency of $10^{4}\;\mathrm{s}^{-1}$ for the second ones, as shown in Table~\ref{table2}. These configurations, except the model with the RMF type EOS, are
already compact enough to have photon or neutrino radii in the equatorial
plane. The values of the neutrino radii together with the equatorial
radii of the stars are given in Table~\ref{table3}. The
radial profile of the ratio in Eq.~(\ref{eq:ddotQdthddotQNdth}) for these
configurations, and of the static case, is presented in Fig.~\ref{fig3}.
\begin{table*}
\centering
\begin{tabular}{|l|l|l|l|l|}
\hline
EOS         & RMF stiff &  STOS 1.0 & Q & CFL 300\\
\hline
$M/R_e$     & 0.2775    & 0.2113    & 0.2682 & 0.2298\\
$R_e/M$     & 3.63      & 4.73      & 3.73   & 4.35\\
$R_e$ [km]  & 14.89     & 19.64     & 15.42  & 18.04\\
\hline
$R_{ph}/M$  & 2.14      & -         & 2.88   & 4.24\\
$R_ph$ [km] & 8.82      & -         & 11.87  & 17.56\\
\hline
\end{tabular}
\caption{The compactness, the equatorial and the (equatorial) photon radii of the compact stars with total mass $M\approx 2.8 M_{\odot}$ and rotational frequency $\Omega\approx 6\times 10^3 \;{\rm s}^{-1}$ and $10^4 \;{\rm s}^{-1}$.}
\label{table3}
\end{table*}

\begin{figure}
\includegraphics[width=8.15cm]{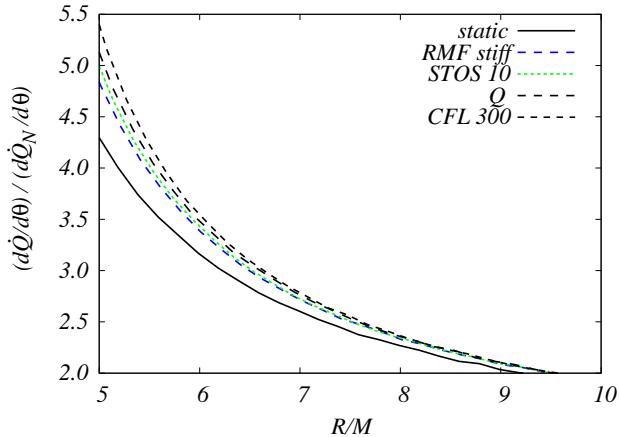}
\centering
\caption{The electron-positron energy deposition rate $\left.\left(d\dot{Q}/d\theta\right)/\left(d\dot{Q}_{N}/d\theta\right)\right|_{\theta=\pi/2}$ in the equatorial plane around rotating neutron and quark stars, with the same total mass $M=2.8M_{\odot}$, and rotational velocities $\Omega =6\times10^3\;{\rm s}^{-1}$ (for STOS and CFL type EOS's), and $\Omega =10^4\;{\rm s}^{-1}$ (for RMF and Q type EOS's). }
\label{fig3}
\end{figure}

The ratios $\left.\left(d\dot{Q}/d\theta\right)/\left(d\dot{Q}_{N}/d\theta\right)\right|_{\theta=\pi/2}$ obtained for the CFL quark stars are somewhat
higher than those obtained for the STOS type EOS's, and this ratio
is essentially 45-50\% higher at the radius $R=5M$, as compared to the same quantity
for the static stars. As for the central objects with RMF and Q
type EOS's, the rotating neutron star produces a deposition rate
about 35\% higher than in the static case. For the quark star at the radius of $R=5M$, the increase in the deposition rate can reach 50\% more than in the static case.
In the region $R\approx5M$, close to the surface of the rotating neutron and quarks stars, and which is relevant
for both the neutron star collapse and the supernova explosion,
there is a considerable
increase in the energy deposition rate of the electron-positron pair
creation, as compared to the Newtonian case.

We have compared not only the neutrino-antineutrino annihilation for different
types of central objects, but we have also studied the effects of the different
rotational velocities of the same central object on the energy deposition
rate. We have chosen the APR EOS for neutron stars, and the Q EOS for quark
stars, to produce sequences of rotating stars with different angular
velocities, but with a fixed total mass of $1.8M_{\odot}$. The rotational
frequencies $\Omega$ are shown in Table~\ref{table3}, together with the other physical
parameters of these configurations.
\begin{table*}
\centering
\begin{tabular}{|l|l|l|l|l|l|l|l|}
\hline
EOS &  APR &  APR  & APR & Q & Q & Q  \\
\hline
$\rho_c\;[10^{15}{\rm g}/{\rm cm}^{3}]$ & 1.28000 & 1.225 & 0.990 & 1.03000 & 0.905000 &  0.960000\\
\hline
$M\;[M_{\odot}]$ & 1.80728 & 1.80 & 1.80504 &  1.80726 & 1.80098 & 1.80266\\
\hline
$M_0\; [M_{\odot}]$ & 2.11215 & 2.11 &  2.03816 & 2.11941 & 2.09223 & 2.02781\\
\hline
$R_e [{\rm km}]$ & 10.9736 & 10.99 &  17.8067 & 1.14149 &  11.9350 & 1.02741\\
\hline
$\Omega [10^3{\rm s}^{-1}]$ & 2.25322 & 5.00 & 7.91730 & 3.14989 & 5.21120 & 12.1428\\
\hline
$\Omega_p [10^3{\rm s}^{-1}]$ & 13.3000 & 13.23 & 6.49228 & 12.5113 & 11.8396 & 14.7565\\
\hline
$T/W [10^{-2}]$  & 0.629016 & 3.41 & 12.6559 & 1.72595 & 5.42560  & 9.2192\\
\hline
$cJ/GM_{\odot}^2$ & 0.487835 & 1.12 &  2.30785 & 0.797423 & 1.42599 &  2.03412\\
\hline
$I [10^{45}{\rm g}\;{\rm cm}^2]$ & 1.90275 & 1.98 & 2.56179 & 2.22488 & 2.40487 & 1.47222\\
\hline
$\Phi_2 [10^{43}{\rm g}\;{\rm cm}^2]$ & 1.48070 & 7.87 & 36.1007 &  4.89741 & 15.9063 & 15.4852\\
\hline
$h_+ [{\rm km}]$ & 3.83236 & 0.00 & 0.00 & 0.00 & 0.00 & 0.00\\
\hline
$h_- [{\rm km}]$ & 6.40902 & -2.11 & 5.85714 & 0.00 & 0.00 & - \\
\hline
$\omega _c/\Omega [10^{-1}]$ & 5.90926 & 5.86 & 5.54520 & 5.36942 &  8.06974 & 8.62910\\
\hline
$r_e  [{\rm km}]$ & 8.07705 & 8.05 & 14.9297 & 8.51829 &  8.99267 & 7.17846\\
\hline
$r_p/r_e$ & 0.980000 & 0.90 & 0.450000 & 0.950000 & 0.850000 & 0.660000\\
\hline
\end{tabular}
\caption{Physical parameters of the neutron stars with APR type EOS and Q type quark stars, with a total mass of  $M\approx 1.8 M_{\odot}$, and with different rotational frequencies.}
\label{table3}
\end{table*}

\begin{figure}
\includegraphics[width=8.15cm]{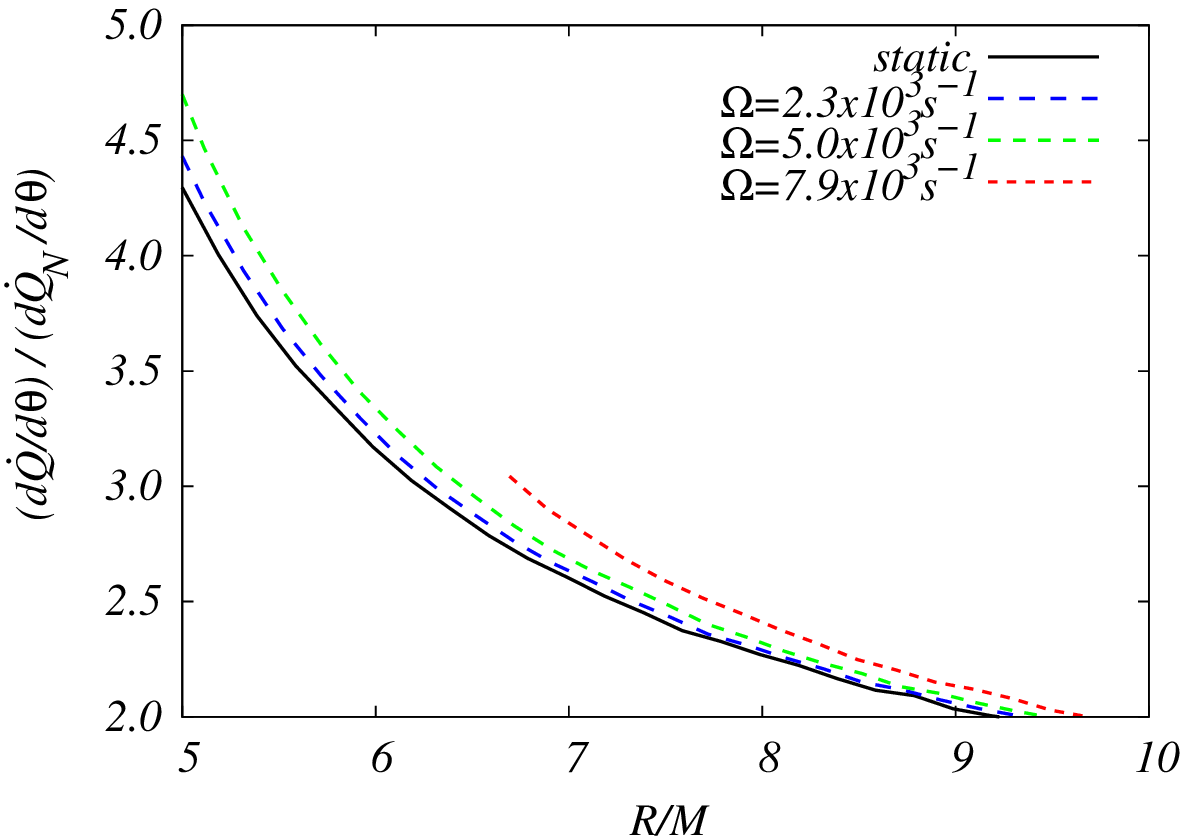}
\includegraphics[width=8.15cm]{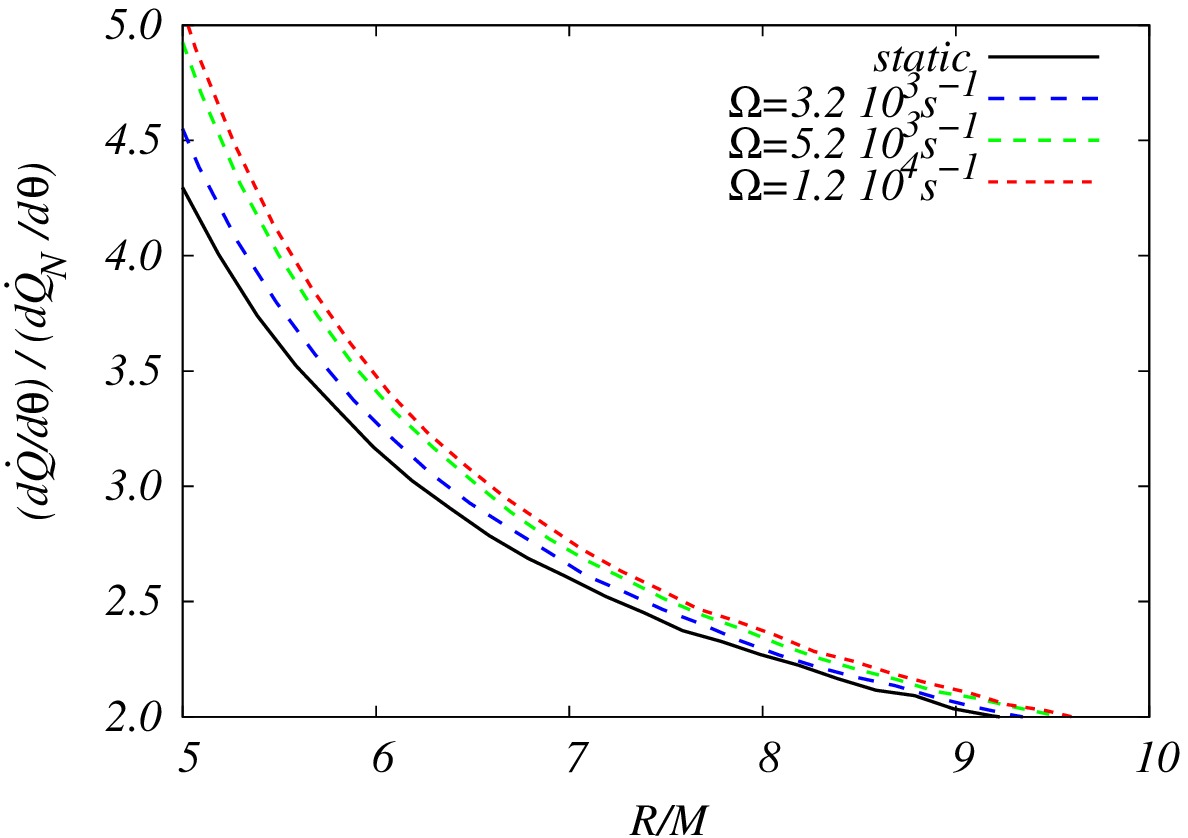}
\centering
\caption{The electron-positron energy deposition rate $\left.\left(d\dot{Q}/d\theta\right)/\left(d\dot{Q}_{N}/d\theta\right)\right|_{\theta=\pi/2}$ in the equatorial plane of a rotating neutron star, with the APR EOS (left hand side), and for a quark star with EOS Q (right hand side), the stars having the same total mass $M=1.8M_{\odot}$, but different rotational velocities. }
\label{fig4}
\end{figure}

In Fig.~\ref{fig4} we present the
ratio $\left.\left(d\dot{Q}/d\theta\right)/\left(d\dot{Q}_{N}/d\theta\right)\right|_{\theta=\pi/2}$ calculated for configurations with increasing angular velocities. With increasing $\Omega$ the
deposition rate is also increasing
for both types of central objects. The ratio $\left.\left(d\dot{Q}/d\theta\right)/\left(d\dot{Q}_{N}/d\theta\right)\right|_{\theta=\pi/2}$
reaches 15\% more for the quark star than its value in the static
case, as the angular frequency approaches the Keplerian limit.
Since the neutron star rotating with the angular velocity of $8\times10^{3}\mathrm{s}^{-1}$
becomes less compact, its equatorial radius is somewhat less than
$7M$, and the deposition rate cannot be integrated to lower radii.
This prevents the star from accumulating much energy released in the
form of electron-positron pairs, even if the ratio given by Eq.~(\ref{eq:ddotQdthddotQNdth})
is very high in the equatorial plane for $R>7M$, as compared to the static case. However, these results show that the rotating ultra-compact objects can convert some of their rotational
energy into the energy of the $e^{+}e^{-}$ pairs.
The faster the central objects rotates, the more energy will be released
in the neutrino-antineutrino pair annihilation process. Then we expect that
ultra-compact stellar object, especially quark stars, rotating close
to the Keplerian limit, can deposit a large amount of energy in their equatorial plane
due to the neutrino-antineutrino annihilation process.

Similarly to the case of the energy deposition, one can define the integrated momentum deposition rate as
\begin{eqnarray}\label{eq:dotP}
\dot{P}(R) & = & 2\int_{0}^{2\pi}\int_{0}^{\pi/2}\int_{R}^{\infty}\dot{p}(r,R,\theta)\sqrt{g_{rr}g_{\theta\theta}g_{\phi\phi}}drd\theta d\phi,
\end{eqnarray}
where $\dot{p}$ is given by Eq.~(\ref{eq:dotp}). In the equatorial plane we introduce the ratio
\begin{equation}\label{eq:ddotPdthddotPNdth}
\left.\frac{d\dot{P}/d\theta}{d\dot{P}_{N}/d\theta}\right|_{\theta=\pi/2}=\frac{\left.\int_{R}^{\infty}\dot{p}(r,R,\theta)\sqrt{g_{rr}(r,\theta)g_{\theta\theta}(r,\theta)g_{\phi\phi}(r,\theta)}dr\right|_{\theta=\pi/2}}{\int_{R}^{\infty}\dot{p}_{N}(r,R)r^{2}dr}
\end{equation}
of the momentum deposition rate in the general relativistic case and of momentum deposition ratio $\dot{P}_N$ corresponding to the Newtonian case. In Fig.~(\ref{fig6})
we present the plots of this quantity as a function of $R/M$ for neutron and quark stars with $1.8M_{\odot}$,  and with a rotation velocity of $5\times10^3 {\rm s}^{-1}$, respectively. This figure is a counterpart of Fig.~(\ref{fig2}), displaying the energy deposition rate for the same configuration of the stellar objects. By comparing the two figures we see that they exhibit the same characteristics for each stars, i. e., the momentum deposition rate is roughly proportional to the the energy deposition rate. Therefore the RMF and STOS EOS type stars produce the highest ratios for the momentum deposition rate.  Quark stars have a smaller momentum deposition rate, and we obtain the smallest values for the neutron stars with DH, APR and BBB type EOSs, respectively.
\begin{figure}
\includegraphics[width=8.15cm]{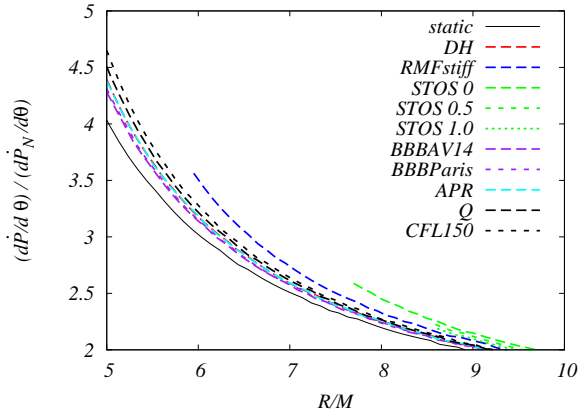}
\centering
\caption{The electron-positron momentum deposition rate $\left.\left(d\dot{P}/d\theta\right)/\left(d\dot{P}_{N}/d\theta\right)\right|_{\theta=\pi/2}$ in the equatorial plane around rotating neutron and quark stars, with the same total mass $M=1.8M_{\odot}$, and same rotational velocity $\Omega =5\times10^3\;{\rm s}^{-1}$.}
\label{fig6}
\end{figure}

\section{Discussions and final remarks}

In the present paper we have considered the general relativistic effects on the energy deposition rate from the neutrino-antineutrino annihilation process near rapidly rotating neutron and quark stars. The energy deposition rate  has been obtained numerically for several equations of state
of the neutron matter, and for two types of quark stars, respectively.  All the general relativistic correction factors, related to this process, can be obtained from the metric of the central compact object. Due to the differences in the space-time structure, the quark stars  present some important differences with respect to the energy deposition rate, as compared to the neutron stars. As a general result we have found that rotation always enhances the energy deposition rate.

A possible astrophysical application of our results could be in the explanation of the physical processes that lead to the formation of the Gamma-Ray Bursts (GRBs). The so-called fireball model
can basically explain the observational facts well, and thus it is strongly favored, and widely accepted today (for recent reviews on GRBs see M\'esz\'aros (2006) and Zhang (2007), respectively). In this
model, the central engine gives birth to some energetic ejecta intermittently, like a geyser, producing a series of
ultra-relativistic shells. The shells collide with each other and produce strong internal shocks. The
highly variable $\gamma$-ray emission in the main burst phase of GRBs should be produced by these internal shocks.
After the main burst phase, the shells merge into one main shell and continue to expand outward.
These merged shells, which moves relativistically, interacts with the surrounding medium to form an external shock. This shock will give a good fraction of its energy to the swept-up electrons and accelerate them to relativistic
velocity. Similarly, a fraction of shock energy will go to the magnetic field. These shocked relativistic electrons move in the magnetic field and emit synchrotron radiation to produce a broadband electromagnetic emission  called "afterglows". According to the Swift observations(see Zhang (2007) for a review), the decay of the X-ray afterglow can be classified roughly into four stages. Stage I (up to $\sim100$ s) indicates the early fast decay phase, usually with flux density decay as $t^{-3}$ to $t^{-5}$; Stage II (up to $\sim 1000$ s) indicates the subsequent
shallow decay phase, with $t^{-0.2}$ to $t^{-0.8}$ ; Stage III (up to $\sim 10^4$ s) is the late normal
decay phase, $t^{-1}$ to $t^{-1.3}$; Stage IV (beyond $10^4$ s) is the late fast decay phase, $t^{-2}$
to $t^{-2.5}$. The first stage is explained as the tail decay of the prompt emission, and the third and the fourth stages can be explained very well in terms of the standard fireball model. Although there are many different explanations of the second stage, the  "shallow decay phase", one of the most popular explanations is the late time continuous energy injection. Obviously, the energy deposition in the equatorial plane cannot be responsible to the prompt gamma-ray emission, which requires a fast and a very narrow beam jet. In fact,  the high energy photons detected by Fermi \citep{Me09} imply that the Lorentz factor of fast jet is over 500.  On the other hand, most deposition energy should come from the equatorial plane. We suggest that the energy deposition along the axis will form a narrow cone jet, which encounters much less material, and hence it can maintain a very large Lorentz factor. On the other hand, the energy deposition in the equatorial plane is higher, but it also encounters much more material, and hence it moves slowly. When the fast jet slows down due to the interstellar medium, this slow but more energetic wind from the equatorial plane can inject additional energy into the fast jet, and result in the shallow decay phase. The detailed calculations will be carried out in our future study.

The neutrino annihilation and the electron-positron pair production also plays an essential role in the astrophysical processes associated with the phase transitions that could take place inside neutron stars.  For example, the sudden phase transition from normal nuclear matter to a mixed phase of quark and nuclear matter induces temperature and density oscillations at the neutrinosphere. Consequently, extremely intense, pulsating neutrino/antineutrino and leptonic pair fluxes will be emitted \cite{Ch09}. During this stage several mass ejecta can be ejected from the stellar surface by the neutrinos and antineutrinos. These ejecta
can be further accelerated to relativistic speeds by the electron/positron pairs, created by the neutrino and antineutrino annihilation outside the stellar surface. In order to produce the Gamma-Ray Bursts, a high neutrino emission rate is necessary. On the other hand, it is important to note that electron-positron pairs can deposit energy much more efficiently than the neutrinos, and the dominant energy deposition process is the neutrino-antineutrino annihilation process \cite{Ch09}. In fact most pairs are created outside the star. A large fraction of the neutrino energy, will be absorbed by the matter very near the stellar surface. When this amount of energy exceeds the gravitational binding energy, some mass near the stellar surface will be ejected, and this mass will be further accelerated by absorbing pairs created from the neutrino and antineutrino annihilation processes outside the star. This process may be a possible mechanism for short Gamma-Ray Bursts \cite{Ch09}.

The neutrino emission rate is also strongly dependent on the temperature. In the standard models of GRBs it is assumed that the central object is surrounded by a degenerate accretion disk, which allows super-Eddington accretion rates, of the order of one solar mass per second \cite{Zh07}. If the central compact object is a neutron or a quark star, such super-Eddington accretion rates can maintain the compact object at very high temperatures, and thus allowing very high neutrino luminosities, as well as a high rate of electron-positron pair production. The neutrino temperature can be estimated by assuming that the accretion power $\eta \dot{M}c^2$, where $\eta $ is the efficiency of the energy conversion and $\dot{M}$ is the accretion rate, is equal to the radiation power $4\pi R^2\sigma T_{\nu }^4$, which gives the temperature as $T_{\nu }=\left(\eta \dot{M}c^2/4\pi R^2\sigma \right)^{1/4}$. By taking $\eta =0.1$, $R=10^6$ cm, and an accretion rate of $\dot{M}=1M_{\odot}/10\;{\rm s}$, we obtain $T_{\nu }=7.14\times 10^{10}$K, a temperature which is of the order of MeV. Therefore, if the accretion disk is fed at a high rate, like, for example, by the fallback material after a supernova explosion, a high neutrino-antineutrino emission rate can be maintained, and this could explain some of the basic properties of the GRB phenomenon.

\section*{Acknowledgments}

We would like to thank to the anonymous referee for suggestions and comments that helped us to significantly improve the manuscript. K. S. C. is supported by the GRF grant number HKU 7011/09P of the government of the Hong Kong SAR. The work of T. H. is supported by the GRF grant number 702507P of the Government of the Hong Kong SAR. Z. K. is indebted to Ritam Mallick for valuable discussions.

\bsp

\label{lastpage}


\begin{thebibliography}{99}

\bibitem[Akmal et al. 1998]{Ak98} Akmal, A., Pandharipande, V. R., \&  Ravenhall, D. G. 1998, Phys. Rev. C, 58, 1804


\bibitem[Alford et al. 1999]{cfl1} Alford, M. G., Rajagopal, K., \& Wilczek, F. 1999, Nucl. Phys. B, 537, 433

\bibitem[Alford et al. 2007]{cfl3} Alford, M. G., Rajagopal, K., Schaefer, T., \&  Schmitt, A. 2008, Rev. Mod. Phys., 80, 1455
\bibitem[Asano \& Fukuyama 2000]{AsFu00} Asano, K. \& Fukuyama, T. 2000, \apj, 531, 949

\bibitem[Asano \& Fukuyama 2001]{AsFu01} Asano, K. \& Fukuyama, T. 2001, \apj, 546, 1019

\bibitem[Asano \& Iwamoto 2002]{AsIw02} Asano, K. \& Iwamoto 2002, \apj, 581, 381

\bibitem[Baldo et al. 1997]{baldo} Baldo, M., Bombaci, I., \& Burgio, G. F. 1997, Astron. Astrophys., 328, 274

\bibitem[Bardeen et al. 1972]{BPT72} Bardeen, J. M., Press, J. H., \& Teukolsky, S. A. 1972, \apj, 178, 347

\bibitem[Baym et al. 1971a]{Ba71a} Baym, G., Bethe, H. A., \& Pethick, C. J. 1971, Nucl. Phys. A, 175, 225

\bibitem[Baym et al. 1971b]{Ba71b} Baym, G., Pethick, C. J., \& Sutherland, P. 1971, \apj, 170, 299

\bibitem[Bhattacharyya et al. 2009]{Ba09} Bhattacharyya, A., Ghosh, S. K., Mallick, R., \& Raha, S. 2009, arXiv0905.3605


\bibitem[Bethe 1990]{Be90} Bethe, H. A. 1990, Rev. Mod. Phys., 62, 801






\bibitem[Bethe \& Wilson 1985]{BeWi85} Bethe, H. A. \& Wilson, J. R., 1985, \apj, 295, 14

\bibitem[Birkl et al. 2007]{Bi07} Birkl, R., Aloy, M. A., Janka, H.-Th., \& Muller, E. 2007, Astron. Astrophys. 463, 51

\bibitem[Bodmer 1971]{Bo71} Bodmer, A. R., 1971, Phys. Rev. D, 4, 1601


\bibitem[Cadeu et al. 2005]{Ca05} Cadeu, C., Leahy D., \& Morsink S. 2005, \apj, 618, 451

\bibitem[Cadeu et al. 2007]{Ca07} Cadeu, C., Morsink S. \& Leahy D. 2007, \apj, 654, 458

\bibitem[Chan et al. 2009]{Cha09}  Chan, T. C., Cheng, K. S., Harko, T., Lau, H. K., Lin, L. M., Suen, W. M., \& Tian,  X. L. 2009, \apj, 695, 732


\bibitem[Cheng et al. 1998]{Ch98} Cheng,  K. S., Dai, Z. G., \& Lu, T. 1998a, Int. J. Mod. Phys. D, 7, 139




\bibitem[Cheng et al. 2009] {Ch09} Cheng, K. S., Harko, T., Huang, Y. F., Lin, L. M., Suen, W. M., \& Tian, X. L. 2009, JCAP, 09, 007


\bibitem[Cooperstein et al. 1986]{Co86} Cooperstein, J., van den Horn, L. J., \& Baron, E. A. 1986, \apj, 309, 653

\bibitem[Cooperstein et al. 1987]{Co87} Cooperstein, J., van den Horn, L. J., \& Baron, E. A. 1987, \apj, 321, L129

\bibitem[Dai et al. 1995]{Da} Dai,  Z. G.,  Peng, Q. H., \& Lu, T. 1995, \apj,  440, 815


\bibitem[Dey et al. 1998]{De98}  Dey, M.,  Bombacci, I.,  Dey, J., Ray, S., \& Samanta, B. C. 1998, Phys.
Lett. B, 438, 123

\bibitem[Douchin \& Haensel 2001]{DoHa01} Douchin, F. \& Haensel, P. 2001, Astron. Astrophys., 380, 151

\bibitem[Feynman et al. 1949]{Fe49} Feynman, R. P.,  Metropolis, N., \& Teller, E. 1949, Phys. Rev., 75, 1561


\bibitem[Goodman et al. 1987]{Go87} Goodman, J., Arnon, D., \& Nussinov S., 1986 \apj, 314, L7









\bibitem[Horvath \& Lugones 2004]{HoLu04} Horvath, J. E., \& Lugones, G. 2004, Astron. Astrophys., 422, L1

\bibitem[Itoh 1970]{It70} Itoh, N. 1970, Prog. Theor. Phys., 44, 291




\bibitem[Kovacs et al. 2009]{Ko09} Kovacs, Z., Cheng, K. S., \& Harko, T. 2009, \mnras, accepted for publication, arXiv:0908.2672

\bibitem[Kubis \& Kutschera 1997]{Kubis} Kubis, S., \& Kutschera, M. 1997, Phys. Lett. B, 399, 191

\bibitem[Landau \& Lifshitz 1998]{LaLi} Landau, L. D. \& Lifshitz, E. M. 1998, The Classical Theory of Fields, Butterworth-Heinemann, Oxford

\bibitem[Lugones \& Horvath 2002]{LuHo02} Lugones, G., \& Horvath, J. E. 2002, Phys. Rev. D, 66, 074017

\bibitem[Mallick \& Majumder 2009]{MaMa09} Mallick, R., \& Majumder, S. 2009, Phys. Rev. D. 79, 023001

\bibitem[Meegan et al. 2009]{Me09} Meegan, C. et al. 2009, Astrophysical J., 702, 791

\bibitem[M\'esz\'aros \& Rees 1992]{MeRe92} M\'esz\'aros, P. \& Rees, M. J. 1992, \mnras, 257, 29

\bibitem[M\'esz\'aros 2006]{Me06} M\'esz\'aros, P. 2006, Rept. Prog. Phys., 69, 2259

\bibitem[Miller et al. 2003]{Mi03} Miller, W. A., George, N. D., Kheyfets, A., \& McGhee, J. M. 2003, \apj, 583, 833





\bibitem[Nozawa et al. 1998]{No98} 	
	Nozawa, T., Stergioulas, N., Gourgoulhon, E., \& Eriguchi, Y. 1998, Astron. Astrophys., 132, 431



\bibitem[Paczynski 1990]{Pa90} Paczynski, B. 1990, \apj, 363, 218

\bibitem[Pandharipande 1971]{Pa71} Pandharipande, V. R. 1971, Nucl. Phys. A, 178, 123



\bibitem[Prasanna \& Goswami 2002]{PrGo02} Prasanna, A. R. \& Goswami, S. 2002, Phys. Lett. B, 526, 27

\bibitem[Rapp et al. 2000]{cfl2} Rapp, R., Schaffer, T., Shuryak, E. V., \& Velkovsky, M. 2000, Ann. Phys. (N. Y.), 280, 35

\bibitem[Ruffert \& Janka 1998]{RuJa98} Ruffert, M. \& Janka, H.-T. 1998, Astron. Astrophys., 338, 535

\bibitem[Ruffert \& Janka 1999]{RuJa99} Ruffert, M. \& Janka, H.-T. 1999, Astron. Astrophys., 344, 573


\bibitem[Salmonson \& Wilson 1999]{SaWi99} Salmonson, J. D. \& Wilson, J. R., 1999, \apj, 517, 895

\bibitem[Salmonson \& Wilson 2001]{SaWi01} Salmonson, J. D. \& Wilson, J. R., 2001, \apj, 561, 950

\bibitem[Shen et al. 1998]{Shen} Shen, H., Toki, H., Oyamatsu, K., \&  Sumiyoshi, K. 1998, Nucl. Phys. A, 637, 435



\bibitem[Stergioulas \& Friedman 1995]{SteFr95} Stergioulas, N., \& Friedman, J. L. 1995, \apj, 444, 306

\bibitem[Stergioulas et al. 1999]{Ste99} Stergioulas, N., Kluzniak, W., \& Bulik, T. 1999, Astron. Astrophys., 352, L116
\bibitem[Stergioulas 2003]{Sterev} Stergioulas, N. 2003, Living Rev. Rel., 6, 3










\bibitem[Witten 1984]{Wi84} Witten, E. 1984, Phys. Rev. D,  30, 272


\bibitem[Zhang 2007]{Zh07} Zhang, B. 2007, Chin. J. Astron. Astrophys., 7, 1

\bibitem[Zhang \& Dai 2009]{ZhDa09} Zhang, D. \& Dai, Z. G. 2009, \apj, 703, 461

\end{thebibliography}
\end{document}